\documentclass{aa}  
\usepackage{siunitx}
\usepackage{wasysym}
\usepackage{makecell}
\usepackage{adjustbox}
\usepackage{booktabs}
\usepackage{natbib}
\usepackage{float}
\bibpunct{(}{)}{;}{a}{}{,}
\usepackage{longtable}
\usepackage{graphicx}
\usepackage{txfonts}
\usepackage[colorlinks=true, linkcolor=blue, citecolor=blue, urlcolor=blue]{hyperref}
\usepackage[version=4]{mhchem}
\usepackage{comment}
\usepackage{xcolor}

\date{}

\usepackage{enumitem}
\setlist[itemize]{label=\textbullet}

\usepackage[table]{xcolor}
\DeclareSIUnit{\MJ}{M_{J}}
\DeclareSIUnit{\RJ}{R_{J}}
\DeclareSIUnit{\ME}{M_\oplus}
\DeclareSIUnit{\Myr}{Myr}
\DeclareSIUnit{\K}{K}
\DeclareSIUnit{\Zp}{Z_{proto}}

\begin{document} 
\title{Evolutionary tracks of giant planets formed by disk instability}
 \titlerunning{Evolutionary tracks of giant planets formed by disk instability}
   \author{Mirco Bussmann\inst{1}\fnmsep\thanks{Corresponding author: mirco.bussmann@uzh.ch} \and Ravit Helled \inst{1}}

   \institute{Department of Astrophysics, University of Zürich, Winterthurerstrasse 190, CH-8057 Zürich, Switzerland
            }

    \abstract
       {The evolution of giant planets depends on their formation history.  The two leading models for giant planet formation are core accretion and disk instability. While core accretion has been extensively studied and several evolutionary models self-consistently connect the planetary origin with the long-term  evolution, comparable models for giant planets formed via disk instability are  absent.}
       {We simulate the evolution of giant planets formed by disk instability and follow their entire evolution including the extended pre-collapse phase, dynamical collapse, and long-term contraction in a unified numerical framework.}
       {The planetary evolution is simulated using the MESPA code  with modifications that allow us to include the extended, low-density gas clumps in the pre-collapse phase. 
       We consider masses between 1 and 12 Jupiter masses and metallicities ranging from 0.5 to 2 times the protosolar value.} 
       {We confirm that the pre-collapse timescale strongly depends on the planetary mass, and that after dynamical collapse the objects reach a state of long-term contraction which lasts for billions of years. We show that metallicity is a major source of uncertainty in mass estimates derived from the age–luminosity relations.  For the metallicity range considered here, we find that for a given measurement of age and luminosity the difference in the inferred mass can be up to 1.5 Jupiter masses.
       We find that our evolution tracks predict masses that are consistent with the measured  dynamical mass constraints for  HR 8799 e, AF Lep b, $\beta$ Pic b and $\beta$ Pic c. We also show that both core accretion and disk instability can lead to very similar long-term evolutionary tracks.}
       {The agreement between our models and dynamical mass measurements suggests that disk instability remains a viable formation pathway for giant exoplanets. The luminosity evolution alone cannot distinguish between the two formation pathways. Finally, we suggest that planetary metallicity must be taken into account when inferring the masses of young giant planets from their luminosities, as it significantly affects their evolution.}

    \maketitle

\nolinenumbers

\section{Introduction}
\label{sec:intro}

The two leading models for giant planets are core accretion \citep{pollackFormationGiantPlanets1996b} and disk instability  \citep{bossGiantPlanetFormation1997a}.
In the core accretion model, giant planets form through a bottom-up growth process.
First, solids accumulate to form a solid/heavy-element core. As the core continues to grow and crosses the critical mass (typically \SIrange[]{5}{15}{\ME}, where \SI{}{\ME} is Earth's mass) it begins to accrete gas rapidly leading to the formation of a gaseous planet \citep[see e.g.,][for a review]{ikomaFormationGiantPlanets2025}.
It is still being investigated whether massive giant planets (several Jupiter masses) can form through this mechanism \citep[e.g.,][]{helledOutstandingQuestionsGiant2026}.
The formation of such massive planets requires a substantial reservoir of gas and they must accrete this material before the protoplanetary disk disperses, a process estimated to occur within only a few Myr \citep{haischjr.DiskFrequenciesLifetimes2001, hernandezSpitzerSpaceTelescope2007, richertCircumstellarDiscLifetimes2018}. This challenge is further compounded by the expectation that growing planets open gaps in the disk, thereby reducing the efficiency of gas accretion \citep{dangeloGrowthJupiterFormation2021}.
A strong argument in favour of core accretion is the observed correlation between stellar metallicity and planet occurrence rate, which is predicted by core accretion formation models \citep{gonzalezStellarMetallicitygiantPlanet1997, santosMetalrichNatureStars2001, idaDeterministicModelPlanetary2004, fischerPlanetMetallicityCorrelation2005, mordasiniExtrasolarPlanetPopulation2012}.
However, a statistical analysis of exoplanets by \citet{santosObservationalEvidenceTwo2017} showed that this correlation weakens for planets with masses above approximately four times Jupiter's  mass (\SI{}{\MJ}), and that for these planets host stars tend to be more metal poor. This suggests the existence of two distinct populations of giant planets, which may indicate that there are two distinct formation mechanisms in play.
Although the transition is unlikely to occur at a sharp mass boundary, lower-mass giant planets may preferentially form through core accretion, whereas higher-mass planets could predominantly originate from disk instability.

In the disk instability model, progenitors of planets, so-called gas clumps, form through instabilities of the protoplanetary disk \citep{toomreGravitationalStabilityDisk1964b}.
These gaseous clumps can have masses on the order of Jupiter masses and are very extended, with initial radii of a few astronomical units (au) \citep[e.g.,][]{nelsonNumericalRequirementsSimulations2006b, forganJeansMassFundamental2011, boleyClumpsOuterDisk2010}. 
Such clumps cool and contract until they eventually collapse and become giant planets \citep{helledEffectsMetallicityGrain2011}.
One of the advantages of the disk instability  mechanism is the short formation timescale  \citep{bossPossibleRapidGas2000, nayakshinTidalDownsizingModel2015}.
Recently, \citet{schibDIPSYNewDisc2025, schibDIPSYNewDisc2025a} presented an updated state-of-the-art population synthesis model for disk instability planets.
In their model, they find that roughly 10\% of protoplanetary disks gravitationally collapse, and about half of them result in a system with at least one companion.
Although they find that most of the formed objects are in the brown dwarf regime, they also find planetary companions, especially at large orbital separations. In addition, some processes of mass-loss could change the predicted masses in this model.
This predicted population of wide-separation, massive companions is consistent with the objects most readily detected by direct imaging surveys \citep[e.g.,][]{bowlerImagingExtrasolarGiant2016}. While the masses of clumps remain uncertain and could be significantly lower \citep[for example][]{dengFormationIntermediatemassPlanets2021}, the disk instability is often used to explain the formation of massive gaseous planets.

Observations offer a possible avenue for distinguishing between different formation pathways, as the luminosity of a young giant planet depends on its thermodynamic state after its formation, described by the initial entropy, and the subsequent thermal evolution. 
The planetary primordial entropy depends on the exact formation model and the local conditions which remain poorly constrained. For simplicity, two end-member scenarios are often being considered:  "hot-start" models where the planet retains high entropy after formation, resulting in a large initial radius and high luminosity \citep[e.g.,][]{baraffeEvolutionaryModelsCool2003a, phillipsNewSetAtmosphere2020}, and "cold-start" models where the energy is assumed to be  efficiently radiated away during formation, leaving the planet in a lower entropy state at the beginning of long-term evolution \citep[e.g.,][]{marleyLuminosityYoungJupiters2007a, fortneyUnifiedTheoryAtmospheres2008}.
Intermediate cases, so-called "warm-start" models have also been explored \citep[e.g.,][]{spiegelSPECTRALPHOTOMETRICDIAGNOSTICS2012, linderEvolutionaryModelsCold2019}. 
Traditionally, cold-start models were associated with core accretion and hot-start models with disk instability.
This picture has become more nuanced as core accretion can also produce warm or hot initial conditions depending on whether accretion energy is efficiently radiated away during formation \citep{mordasiniCharacterizationExoplanetsTheir2012, cummingPrimordialEntropyJupiter2018}.
Conversely, in disk instability, the initial gas clump forms through a self-gravitational binding process and therefore there is no obvious mechanism that removes entropy from the gas, thus, naturally favouring a hot-start \citep[e.g.,][]{galvagniCollapseProtoplanetaryClumps2012a}. 
Currently, no evolutionary framework directly linked to the disk instability formation scenario exists. 
With the growing number of directly imaged planets with dynamical mass constraints, it is becoming possible to test whether evolutionary models, including their initial entropies, are consistent with observations.
It is therefore desirable to establish a direct link between formation models and planetary evolutionary tracks, enabling a self-consistent comparison with observation and testing existing formation models. 

In this work, we simulate the long-term evolution of giant planets that form via disk instability.
We follow the thermal evolution of gravitationally bound gas clumps from the pre-collapse stage, through dynamical collapse and subsequent post-collapse evolution up to ages of several billion years. Our models cover a range of planetary masses and compositions.
The paper is structured as follows. In Section \ref{sec:methods}, we describe how the initial gas clump models are generated and how the different evolutionary stages are modelled. In Sections \ref{sec:full_evol} and \ref{sec:evol_metal}, we present the full evolution and investigate its sensitivity to composition. In Section \ref{sec:comparison}, we compare our evolutionary tracks with traditional hot-start models and with models linked to core accretion. In Section \ref{sec:obs_with_dyn}, we use our cooling tracks to infer the masses of directly imaged planets and compare them with their measured dynamical masses. Finally, we summarize our results and conclusions in Section \ref{sec:summaryandconclusion}.

\section{Methods}
\label{sec:methods}

We simulate the evolution of gaseous clumps  by solving the stellar structure equations \citep[e.g.,][]{kippenhahnStellarStructureEvolution2012} using the Henyey method \citep{henyeyNewMethodAutomatic1964}.
We use the Modules for Experiments in Stellar Astrophysics (MESA) code (\citealt{paxtonModulesExperimentsStellar2011, paxtonModulesExperimentsStellar2013, paxtonModulesExperimentsStellar2015, paxtonModulesExperimentsStellar2018, paxtonModulesExperimentsStellar2019, jermynModulesExperimentsStellar2023}), which was initially built for stars and simple planet models and therefore required some modifications to make it adequate to the different evolutionary stages.
In the pre-collapse stage, the gas clumps are very cold and reach temperatures well below \SI{100}{\K}, which is the lower threshold for even the coldest opacity tables available by default in MESA.
We therefore implemented a routine that uses the Rosseland mean opacities from \citet{pollackCalculationRosselandMean1985} and \citet{alexanderLowtemperatureRosselandOpacities1994} where the grains are assumed to be in chemical equilibrium with a gas of solar composition and to have an interstellar grain size distribution. These opacities are calculated down to \SI{10}{K}. We note that updated calculations of opacities in low temperatures and pressures would be valuable.
We also include a linear scaling of the opacity with metallicity to account for the larger amount of grains, similar to \citet{helledEffectsMetallicityGrain2011}.
For the post-collapse evolution, we used standard opacity tables in MESA which are based on the work of \citet{freedmanLineMeanOpacities2008} and privately communicated by R.~S.~Freedman in 2011 for the low-temperature regime.
In the high-temperature regime, we use the default values of MESA and for the deep interior of giant planets electron conduction is expected to be significantly more efficient, which is described by an extended version of the tables based on \citet{cassisiUpdatedElectronConductionOpacities2007}.

For both evolutionary stages, we use Equations of State (EoS) built with the python module \textsc{tinyeos\footnote{https://github.com/tiny-hippo/tinyeos}}, which uses the additive volume law to combine EoS of different substances, and implement them using the MESA extension MESPA\footnote{https://github.com/uzhplanets/mespa} \citep[Modules for Experiments in Stellar and Planetary Astrophysics;][]{helledGiantPlanetEvolution2025}. This extension makes MESA a state-of-the-art planetary evolution code and includes the most widely used EoS for planetary evolution.
Due to the very low temperatures (and pressures) of the clumps during pre-collapse, the objects are modelled using the SCvH EoS \citep{saumonEquationStateLowMass1995a} for hydrogen and helium, with the heavy elements being represented by a 50/50 mixture of \ce{H2O} and \ce{SiO2} using QEoS \citep{moreNewQuotidianEquation1988, vazanEffectCompositionEvolution2013a}. 
After the collapse, we use the same EoS for the heavy elements, but the more appropriate CMS EoS \citep{chabrierNewEquationState2019, chabrierNewEquationState2021} for hydrogen and helium, which is often used in recent planetary evolution simulations \citep[e.g.,][]{howardEvolutionJupiterSaturn2024, eberleinEvolutionInternalStructure2025}. 
The transition from the SCvH EoS to the CMS EoS during the dynamical collapse introduces a slight inconsistency. Depending on the choice of the EoS, for the same initial parameters (i.e., mass, composition and initial entropy) the initial post-collapse radius can vary by up to 50\%. Although this is a significant difference in the initial configuration shortly after the collapse, this only leads to a short-lived effect that dissipates after \SI{1}{\Myr} and therefore does not affect the long-term evolution. At the same time,  the long-term evolution is significantly altered depending on which EoS is used.  The CMS hydrogen-helium EoS is more appropriate for the pressure-temperature regime associated with the long-term evolution and we therefore use this EoS for this later stage. Note that it is not possible to model the pre-collapse phase with the CMS EoS as it does not include the physical conditions of clumps (very low temperatures and densities).
More details on the importance of the EoS and its effect on the evolution can be found in Appendix \ref{app:importance_of_eos}. 

The initial model was guided by the simulation of a \SI{12}{\MJ} clump model presented in \citet{humphriesConstrainingInitialPlanetary2019b}.
Using built-in controls in MESA, we relax its mass, composition, and total entropy to reach the desired initial condition.
We consider masses between 1 and \SI{12}{\MJ}, which are typical masses inferred by  hydrodynamical simulations \citep[e.g.,][]{mayerEvolutionGravitationallyUnstable2004, boleyClumpsOuterDisk2010}.
As gas clumps are expected to form early, their composition is often assumed to be similar to that of the host star. 
However, due to the formation of spirals in the disk, which are the region where clumps typically form, there might be an enhancement of metals that  could lead to higher metallicities \citep{boleyPOSSIBILITYENRICHMENTDIFFERENTIATION2010}.

To investigate the effect of metallicity on the evolution we use metallicities of 0.5, 1 and 2 times the protostellar metallicity of $ Z_{proto} = 0.0187$ \citep{loddersSolarSystemElemental2025}.
The hydrogen and helium mass fractions are adapted such that their ratio stays constant.
Using the solar abundances from \citet{asplundSolarChemicalComposition2006} this translates to $[Fe/H] = -0.151$, $+0.151$ and $+0.461$, respectively.\\

The thermal evolution of the pre-collapse stage is followed until the central temperature reaches $\sim$\SI{2000}{K}, when the dynamical collapse is initiated.
The dynamical collapse is not directly modelled, instead, we assumed an   adiabatic collapse \citep{rakavyOxygenBurningStarsPreSupernova1967, kovetzNewEfficientStellar2009, vazanEVOLUTIONSURVIVALPROTOPLANETS2012a}, and construct a new post-collapse model in which the radius is selected so that the total entropy of the object is the same as before collapse. 
 The post-collapse radius is determined by interpolating a pre-generated table that maps mass, composition, and total entropy to radius. 
After collapse, the long-term evolution is simulated until the planets reach ages of several gigayears.
\\

\section{The Planetary Evolution}
\label{sec:full_evol}
\begin{figure}[h]
    \centering
    \includegraphics[width=1\linewidth]{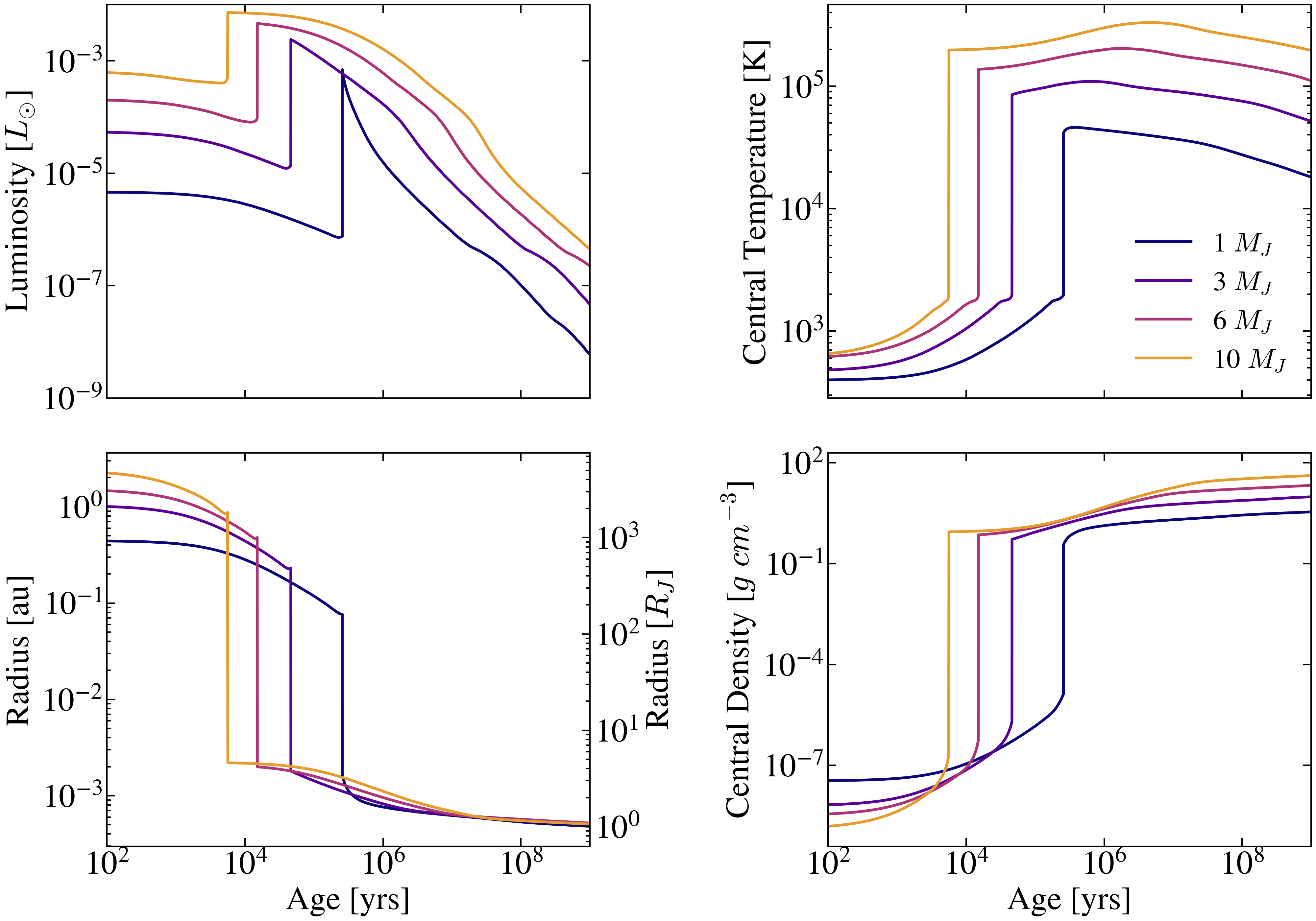}
    \caption{Evolutionary tracks for a 1, 3, 6 and \SI{10}{\MJ} clump, including the pre-collapse, the dynamical collapse and the long-term post-collapse evolution. Top left: Luminosity evolution in solar units. Top right: Central temperature in Kelvin. Bottom left: Radius evolution in au (left axis) and \SI{}{\RJ} (right axis). Bottom right: Central density in $g$ $cm^{-3}$. The sharp transitions in each panel corresponds to the dynamical collapse.}
    \label{fig:full_evolution_1_3_5_10MJ}
\end{figure}

Figure \ref{fig:full_evolution_1_3_5_10MJ} shows the full evolutionary tracks for objects with 1, 3, 6 and \SI{10}{\MJ}. 
The evolution can be split into three phases: pre-collapse,  dynamical collapse, and long-term (post-collapse) evolution.
During the pre-collapse phase the gas clumps are still very extended with  radii of the order of astronomical units. 
The gas clumps slowly contract, leading to a steady increase in the central density and temperature.
Due to the decreasing surface area the luminosity slowly decreases as the gas clump evolves.

Note that the clumps' luminosities are very high despite the low surface temperatures due to their very large radii. The surface temperature remains below 100 K throughout the pre-collapse phase making the detection of such objects extremely challenging. 
The pre-collapse timescale strongly depends on the planetary mass:  while the pre-collapse for a \SI{1}{\MJ} gas clump is $\sim 2.6 \times 10^5$ yr, it is $\sim 5.6 \times 10^3$ yr for a  clump with a mass of \SI{10}{\MJ}. Our results are consistent with previous studies that find similar trends \citep[e.g.,][]{helledCoreFormationGiant2008a, vazanEVOLUTIONSURVIVALPROTOPLANETS2012a, galvagniCollapseProtoplanetaryClumps2012a}. 
The shorter pre-collapse stage of more massive objects is driven by the greater compression, which leads to higher temperatures and substantially increased luminosities. This trend is reflected in the Kelvin-Helmholtz timescale, 

\begin{align}
    t_{KH} &= \frac{E_g}{L} = \frac{GM^2}{RL},
\end{align}
where $E_g$ is the gravitational binding energy, $G$ the gravitational constant, and $M$, $R$, and $L$ are the mass, radius and luminosity of the planet, respectively. The Kelvin-Helmholtz timescale provides a good estimate of the duration of the pre-collapse phase. 
During most of the pre-collapse phase the clump is mostly convective, only in the last stage shortly before the collapse occurs the opacity drops in the centre which leads to a radiative core.
The pre-collapse phase ends once the central temperatures reach $\sim$2000 K, when molecular hydrogen starts to dissociate. The exact temperature at which this occurs depends on the mass of the clump, as this changes the central density and pressure and therefore also the exact dissociation temperature. For less massive clumps (around \SI{1}{\MJ}) the dissociation temperature is slightly above \SI{2000}{\K} while for more massive clumps (around \SI{10}{\MJ}) it is somewhat lower. As the clump continues to contract and enters the dissociation regime, the energy released by compression is largely consumed by the dissociation process. As a result, while the central density can increase by orders of magnitude the temperature remains nearly constant and the clump's centre loses its pressure support which leads to collapse. This collapse, also known as ``dynamical collapse",  occurs  very quickly, roughly on the order of the free-fall timescale of a few years.
After the collapse, the protoplanet reaches a few Jupiter radii (\SI{}{\RJ}), in our case between \SIrange{2}{3.5}{\RJ} and is again in hydrostatic equilibrium. 
The exact radius depends on the mass of the clump, its composition, and entropy. 

\begin{figure}[h]
    \centering
    \includegraphics[width=1\linewidth]{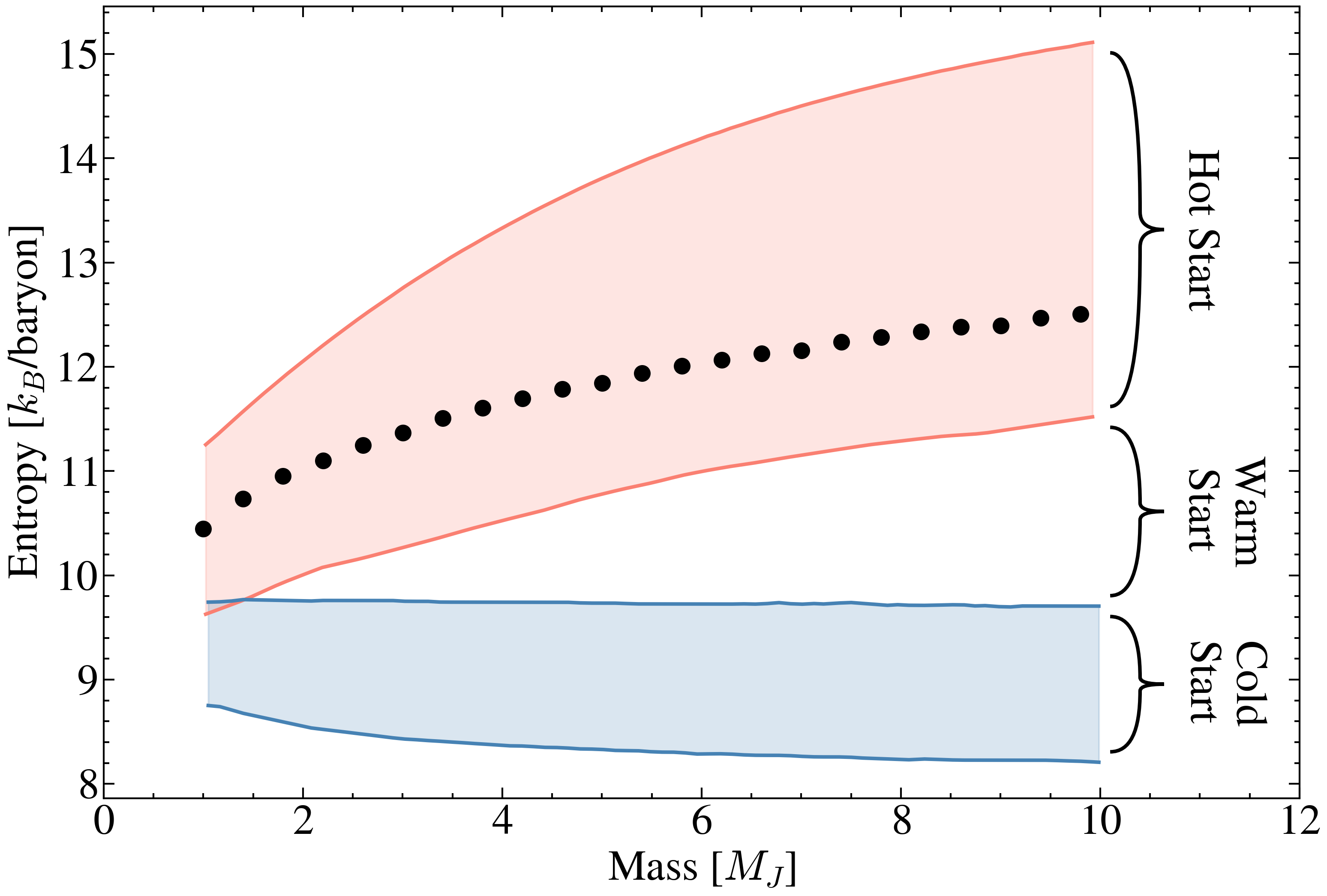}
    \caption{Initial entropy of our models post dynamical collapse for the different masses. The entropy is determined by following the evolution of gas clumps and under the assumption of an adiabatic collapse. Also shown are different regimes of traditional cold, warm, and hot starts \citep{spiegelSPECTRALPHOTOMETRICDIAGNOSTICS2012}.}
    \label{fig:initial_entropy}
\end{figure}

Figure \ref{fig:initial_entropy} shows the initial post-collapse entropies we find for the different protoplanetary masses.
As expected, these are very high initial entropies and according to the characterization in e.g., \citet{spiegelSPECTRALPHOTOMETRICDIAGNOSTICS2012}, they are similar to classical hot-start models (see Section \ref{sec:comparison}).
At that point, the long-term evolution begins. The planet then slowly contracts and cools for billions of years. The contraction and cooling are governed by the ability of the planet to radiate its primordial energy. As a result, the cooling strongly depends on the atmospheric opacities. 
At the beginning of the long-term evolution the central temperature increases.
This is due to an initial rapid contraction of the protoplanet which leads to a faster increase in the central density and an increase in the central temperature.

\begin{figure}[h]
    \centering
    \includegraphics[width=1\linewidth]{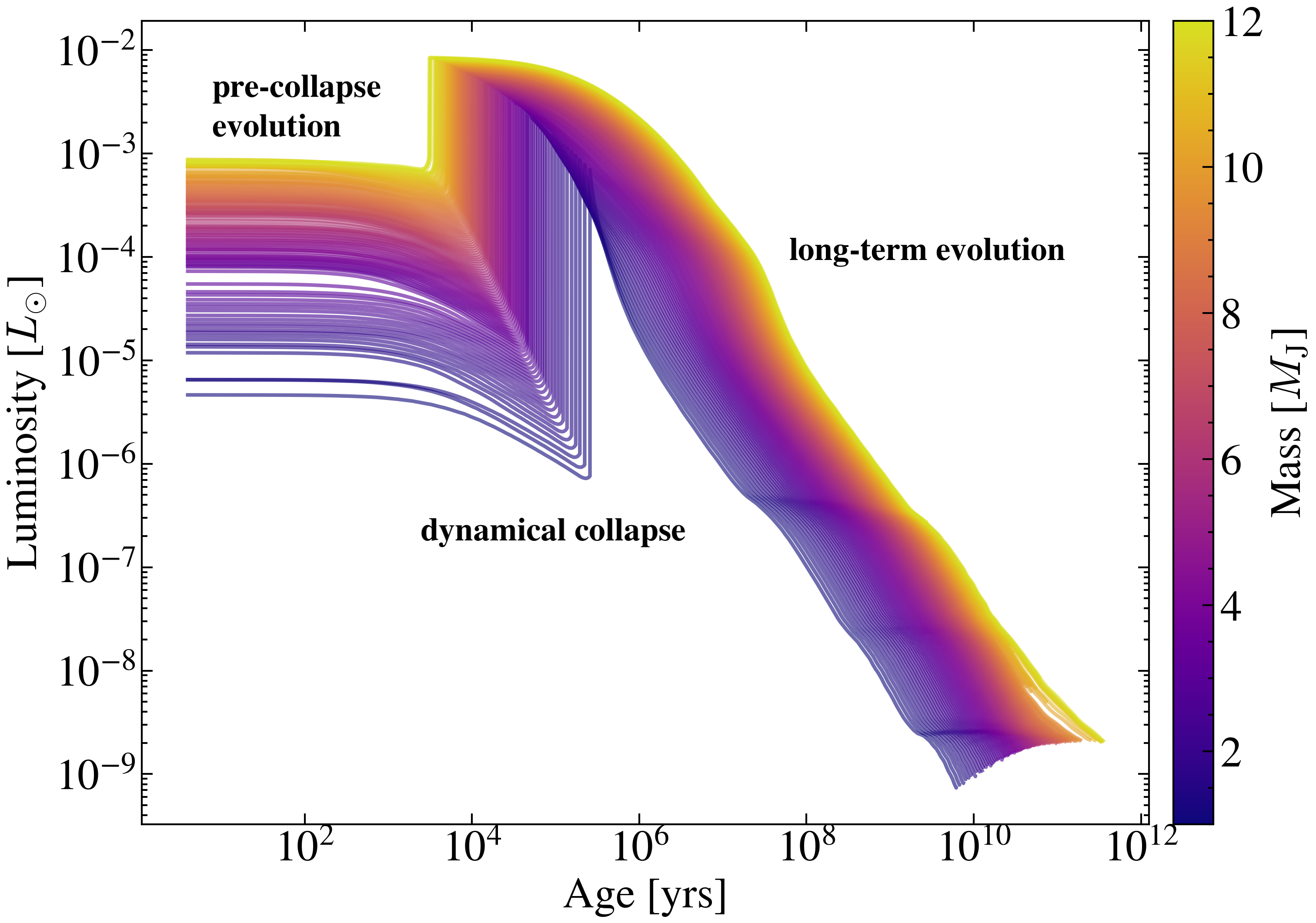}
    \caption{Complete cooling curves for different planetary masses. The evolution includes the pre-collapse evolution, the dynamical collapse and the long-term contraction.}
    \label{fig:cooling_curves}
\end{figure}

Only once the contraction slows down, the central temperature cools down as well. 
The luminosity curve for the \SI{1}{\MJ} clump has a spike because the plots are in log-log scale. 
After the \SI{1}{\MJ} clump has collapsed it possesses a higher luminosity than the \SI{3}{\MJ} clump because the \SI{3}{\MJ} clump has collapsed much earlier. Thus, at the point the \SI{1}{\MJ} gas clump collapses, the \SI{3}{\MJ} clump has already cooled and contracted, which leads to lower luminosities at that time.
Figure \ref{fig:cooling_curves} shows the cooling curves for the protoplanets for masses between 1 and \SI{12}{\MJ} for a proto-solar composition.
Note that the starting time of the evolution, $t_0$, corresponds to the time at which the gas clump forms in the disk. However, as gaseous clumps are expected to form rather early on in the disk, it is reasonable to treat the time of the cooling curves as the age of the system.

\section{The importance of metallicity}
\label{sec:evol_metal}
\begin{figure}[h]
    \centering
    \includegraphics[width=1\linewidth]{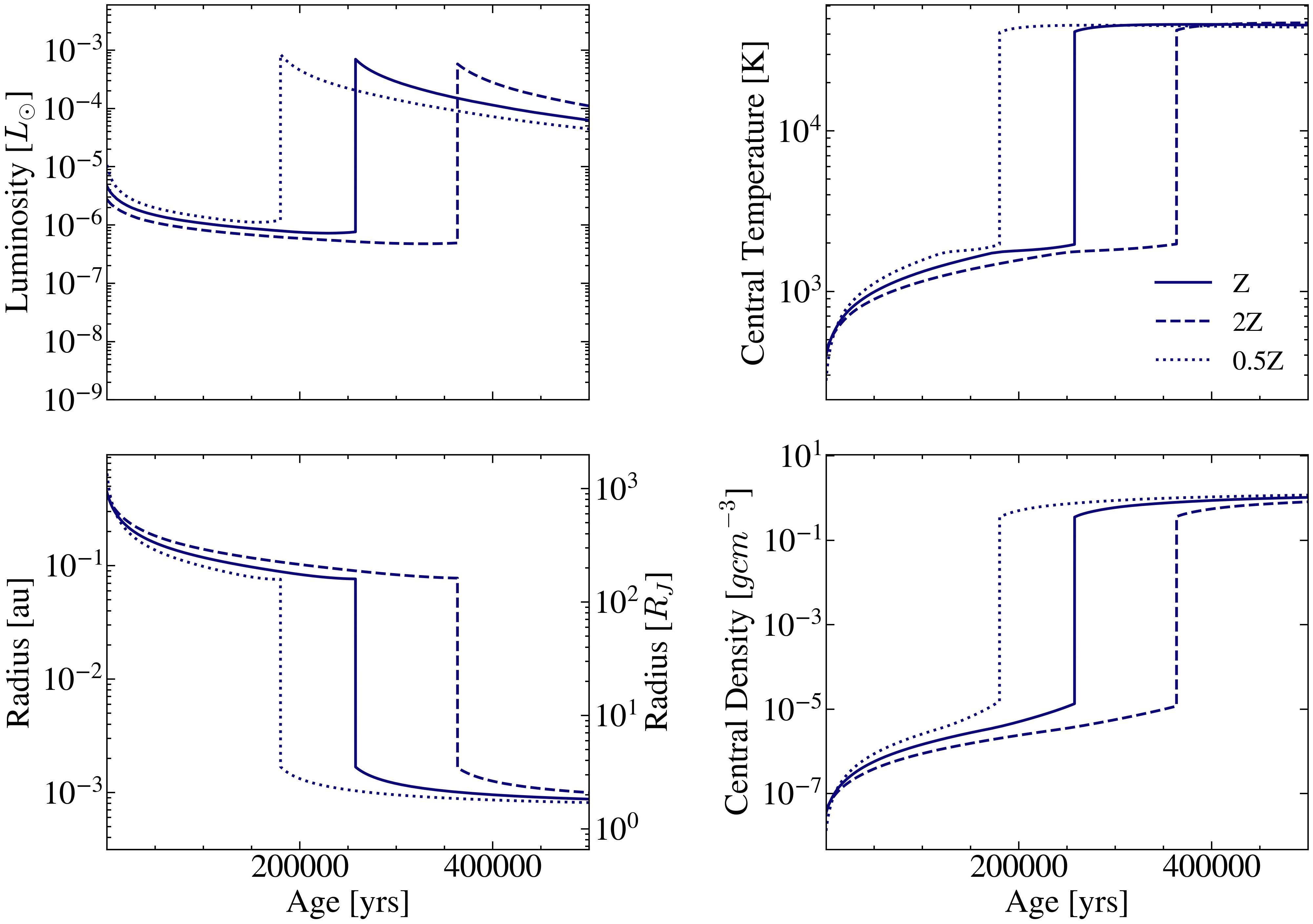}
    \caption{Evolution of luminosity, central temperature, radius and central density throughout the pre-collapse phase for a \SI{1}{\MJ} gas clump. Higher metallicities lead to higher opacities and thus prolong the pre-collapse phase.}
    \label{fig:precollapse_diff_Z_evol}
\end{figure}

Different metallicities in gaseous clumps significantly alter the thermal evolution.  
Figure \ref{fig:precollapse_diff_Z_evol} shows the pre-collapse evolution of a \SI{1}{\MJ} gas clump for three different assumed  metallicities. 
We find that the main effect of metallicity during the pre-collapse stage is caused by the opacity dependence on metallicity. 
For a \SI{1}{\MJ} gas clump, the pre-collapse timescale in the \SI{2}{\Zp} case is nearly twice that of the \SI{0.5}{\Zp}.
We find this relation for all masses, which is in good agreement with \citet{helledEffectsMetallicityGrain2011} who found that the pre-collapse timescale is proportional to the metallicity of the gas clump.
This strong dependence arises because the opacity in the pre-collapse  phase is dominated by dust grains, whose contribution scales approximately linearly with metallicity. As a result, a higher metallicity significantly increases the opacity, reducing the cooling efficiency and prolonging the contraction timescale. Of course if there is a different scaling the relation would change accordingly.

The assumed metallicity influences the cooling curves in three main ways: 
First, an increased metallicity prolongs the pre-collapse phase and delays the onset of the dynamical collapse. As a result, the post-collapse evolution of metal-rich protoplanets begins at a later time in comparison to metal-poor protoplanets.
Second, after dynamical collapse, a higher metallicity leads to a more compact configurations. Consequently, metal-rich protoplanets begin their long-term evolution from a more contracted state, which corresponds to a lower initial luminosity. Finally, the metallicity also determines the opacity and thus the cooling efficiency which dictates the cooling curves on longer timescales. 

\begin{figure}[h]
    \centering
    \includegraphics[width=1\linewidth]{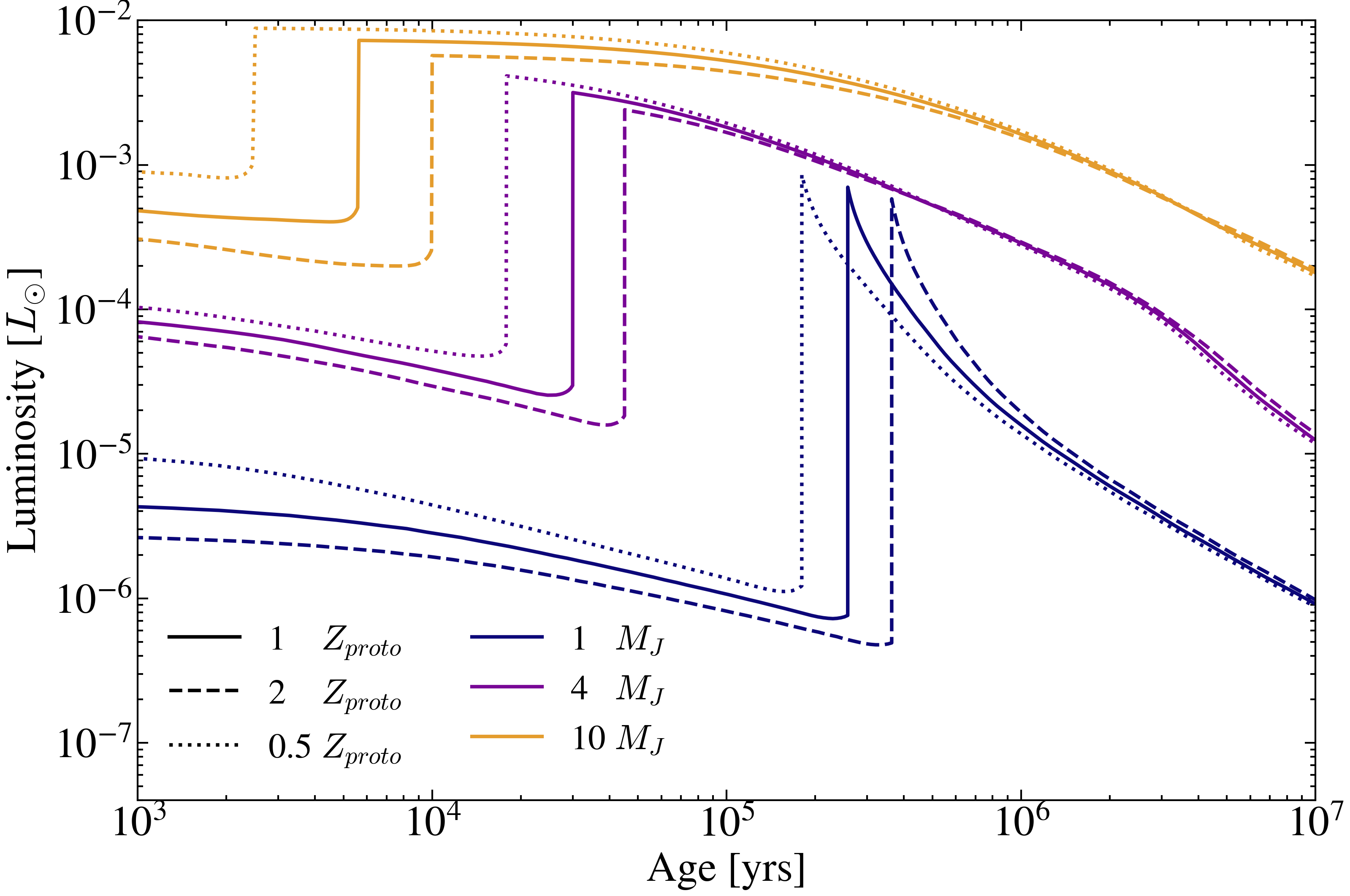}
    \caption{Luminosity evolution for protoplanets with masses of  1, 4 and \SI{10}{\MJ} assuming three different metallicities. The plot shows the end of the pre-collapse phase, the dynamical collapse and the initial phase of the post-collapse, long-term evolution.}
    \label{fig:cooling_curves_metallicity_IC}
\end{figure}

For massive objects, the pre-collapse phase is typically short, and therefore differences in its duration have only minor effects on the subsequent evolution. As a result, the largest differences shortly after the dynamical collapse arise from the initial structure of the protoplanet. Protoplanets with higher metallicities are more compact and therefore have lower luminosities shortly after collapse (see Figure \ref{fig:cooling_curves_metallicity_IC}). As protoplanets evolve, the higher opacity associated with higher metallicity slows their cooling and contraction. Consequently, their luminosity decreases more slowly than that of lower-metallicity protoplanets. This eventually leads to a reversal in the luminosity trend, with the metal-rich protoplanets becoming more luminous. For example, for a \SI{10}{\MJ} protoplanet, this transition occurs at approximately \SI{4}{Myr}. For lower-mass protoplanets, it occurs at earlies times. For the lowest-mass protoplanets, the duration of the pre-collapse phase becomes important as it becomes comparable to the evolutionary timescale after collapse. Typical pre-collapse timescales are of the order $10^5$ years, so a factor of two difference can significantly delay the pre-collapse phase. As a result, when the metal-rich protoplanet finally collapses, the metal-poor protoplanet has already had time to contract and cool. The combination of this evolutionary delay and the slower cooling caused by the higher opacity means that the metal-rich protoplanet remains more luminous throughout its post-collapse evolution. This is of course assuming that opacity scales with metallicity which may not be the case if grain growth and settling are considered \citep{helledEffectsMetallicityGrain2011}.

\begin{figure}[h]
    \centering
    \includegraphics[width=\hsize]{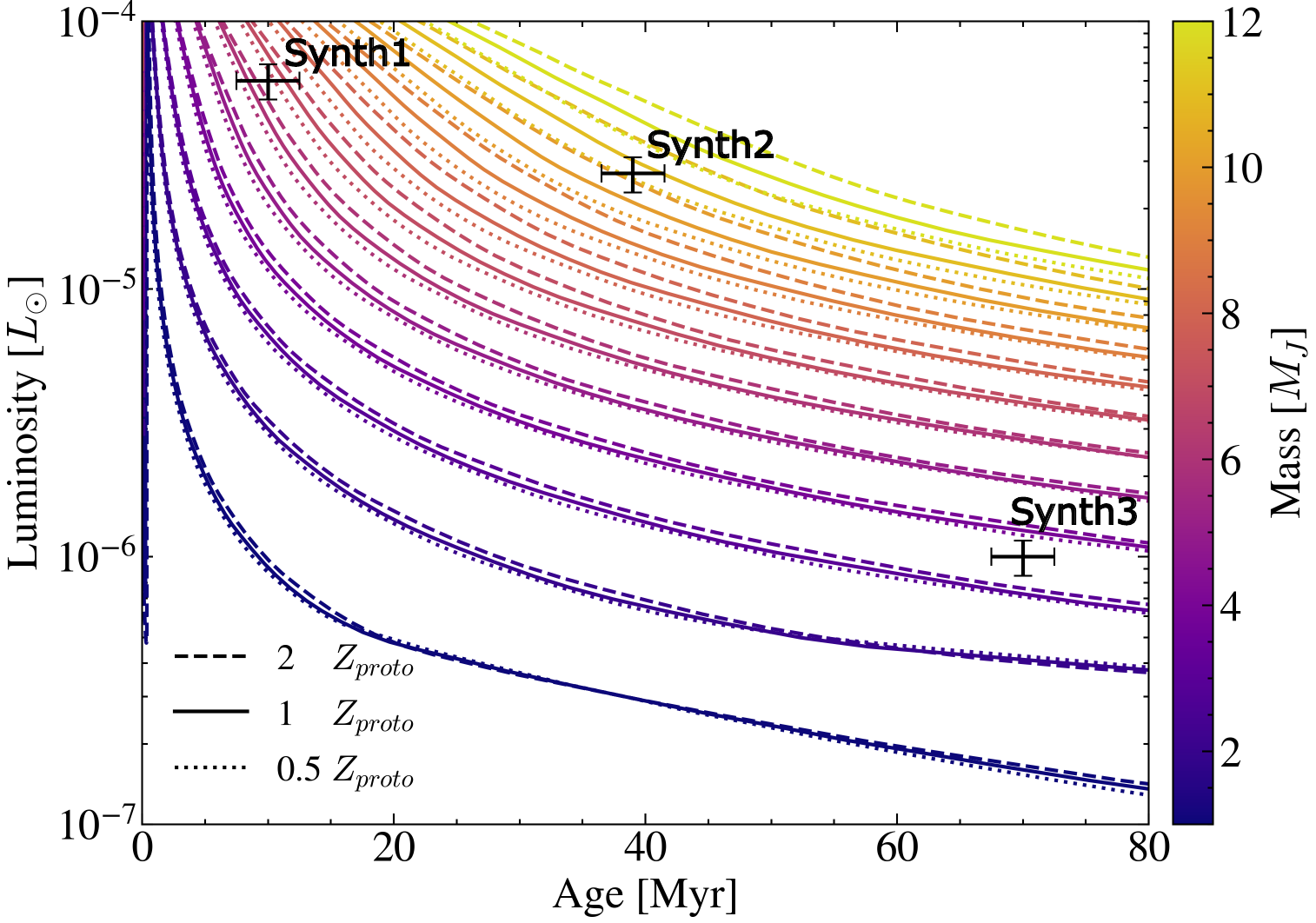}
    \caption{Cooling curves of protoplanets for different assumed masses and metallicities. The plotted ages correspond to typical ages of systems with directly imaged planets. The plotted error bars correspond to synthetic observations as listed in Table \ref{tab:synth_obs}. We use these synthetic measurements to  determine the inferred masses at different conditions (see text for details).}
    \label{fig:cooling_curves_metallicity}
\end{figure}

For all the masses we considered, we find that the evolution after the first few Myr is governed primarily by the opacity. Figure \ref{fig:cooling_curves_metallicity} presents the luminosity evolution of protoplanets with different masses and metallicities over $8 \times 10^7$ yr, an age corresponding to estimated ages of directly imaged planets. Overall, we find that metallicity can have a significant impact on the luminosity evolution, particularly for higher-mass protoplanets. The importance of metallicity, however, varies with age. At very young ages (a few Myr) and at later times ($\gtrsim 70$ Myr), the differences between models with different metallicities remain relatively modest. In contrast, metallicity effects are strongest at intermediate ages, which coincide with the age range of most directly imaged planets. To illustrate how metallicity affects the inferred mass from luminosity measurements, we include three synthetic observations in Figure \ref{fig:cooling_curves_metallicity}. We also list the corresponding inferred masses with uncertainties for the different compositions in Table \ref{tab:synth_obs}. The first synthetic observation, \textit{Synth1}, represents a very young planet. At such early ages, the cooling tracks of different masses are closely clustered, resulting in large uncertainties in the inferred mass. For each metallicity considered separately, the uncertainty exceeds  \SI{1}{\MJ}. The variation in the inferred mass between different metallicities is smaller and corresponds to about \SI{0.5}{\MJ}. The second synthetic observation, \textit{Synth2}, represents a massive planet at an intermediate age, characteristic of many currently known directly imaged planets. At this stage, the cooling tracks associated with different masses are more separated, reducing the uncertainty in the inferred mass to $\sim$  \SI{0.38}{\MJ}. However, the unknown metallicity introduces a strong degeneracy in this regime, with the inferred mass for the same luminosity and age varying by more than \SI{1.5}{\MJ} across the metallicities considered. The final synthetic observation, \textit{Synth3}, corresponds to an older, lower-mass planet. In this part of parameter space, the inferred mass is rather insensitive to the assumed metallicity and the resulting mass estimates remain largely unchanged across the different compositions considered. These examples highlight that the planetary composition (and opacity) can  substantially bias the mass estimates of directly imaged planets, particularly for the massive planets at ages of a few tens of Myr.

\begin{table*}[ht]
\centering
\setlength{\tabcolsep}{12pt} 
    {\renewcommand{\arraystretch}{1.3}
    \caption{Synthetic observations with inferred masses.}
    \begin{tabular}{cccccc}
    \toprule \toprule
        Planet              & Age [\SI{}{\Myr}] & log($L$/$L_\odot$) & $M_{evol}^{2 Z_{proto}}$ [\SI{}{\MJ}] & $M_{evol}^{1 Z_{proto}}$ [\SI{}{\MJ} ]& $M_{evol}^{0.5 Z_{proto}}$ [\SI{}{\MJ}] \\ \midrule
        \textbf{Synth1}     & $10_{-2.5}^{+2.5}$   & $-4.22_{-0.07}^{+0.06}$ & 6.27 $\pm$ 1.08 & 6.52 $\pm$ 1.12 & 6.78 $\pm$ 1.18\\
        \textbf{Synth2}     & $40_{-2.5}^{+2.5}$   & $-4.60_{-0.07}^{+0.06}$  & 9.73 $\pm$ 0.38  & 10.17 $\pm$ 0.38 & 11.33 $\pm$ 0.47\\
        \textbf{Synth3}     & $70_{-2.5}^{+2.5}$   & $-6.00_{-0.07}^{+0.06}$  & 1.70 $\pm$ 0.02 & 1.66 $\pm$ 0.16 & 1.72 $\pm$ 0.17 \\
    \bottomrule
    \bottomrule
    \end{tabular}
    }
    \tablefoot{This table describes synthetic observations with inferred masses to illustrate the degeneracy due to metallicity for different ages and masses.}
    \label{tab:synth_obs}
\end{table*}

{\section{Comparison with previous studies}
\label{sec:comparison}

In this section, we compare our evolutionary tracks from two perspectives. First, we compare them with the state-of-the-art hot-start models \textsc{atmo 2020} \citep{phillipsNewSetAtmosphere2020} and \textsc{sonora bobcat} \citep{marleySonoraBrownDwarf2021a} to compare our  predictions with previous studies that model the planetary evolution  assuming a hot-start. Second, we compare our evolution tracks with evolution models that correspond to formation via core accretion. This allows us to investigate whether the different formation histories leave observable signatures in the long-term evolution of giant planets.

Since the initial entropies of our models at the beginning of the long-term evolution are comparable to those typically adopted in hot-start calculations (see Figure \ref{fig:initial_entropy}) it is possible to compare them and expect similar results. It is important to note, however, that a direct comparison is challenging: 
Although all the three models correspond to hot-start scenarios, our evolutionary tracks also include the preceding pre-collapse phase of gaseous clumps. 
The self-consistent treatment from the pre-collapse stage through to the long-term evolution means that both the initial configuration and the starting time for the long-term evolution are not arbitrarily chosen. In contrast, the initial conditions for \textsc{atmo 2020} and \textsc{sonora bobcat} are set at the beginning of the hot-start phase, which may not correspond to the same evolutionary time as in our models, resulting in a shift.
Figure \ref{fig:cooling_curves_comparison_other_studies} shows the luminosity evolution for four masses from the three models at ages relevant for directly imaged planets. 

Overall, we find that our model is in good agreement with previous models across the entire mass and age range. 
All the models reproduce a similar cooling behaviour characterized by rapid cooling and contraction during the first few Myr followed by a more moderate contraction.  
For the lowest-mass objects, the agreement between the different evolutionary calculations is excellent. Systematic differences become increasingly apparent for higher masses, where our models generally predict slightly lower luminosities at a given age. Apart from differences inherent to our coupled treatment of formation and evolution, these differences are mainly due to a different treatment of the atmospheric boundary conditions. While our calculations employ a grey atmosphere approximation combined with opacity tables, \textsc{atmo 2020} and \textsc{sonora bobcat} use more sophisticated non-grey atmosphere models.
Nevertheless, the overall agreement is reassuring and demonstrates that our self-consistent approach from pre-collapse to long-term evolution produces cooling tracks that are robust and reliable for characterizing directly imaged exoplanets. Finally, we note that our atmospheric model is very simple. In reality, processes such as atmospheric chemistry, condensation, and clouds formation could affect the planetary contraction and the luminosities (especially at young ages, see \citet{morleySonoraSubstellarAtmosphere2024}). To demonstrate the importance of this effect a comparison between our models and those inferred by  \textsc{sonora red diamondback} \citep{ davisSonoraSubstellarAtmosphere2025} are presented in Appendix \ref{sec:comparison_diamondback}.

\begin{figure}[h]
    \centering
    \includegraphics[width=\hsize]{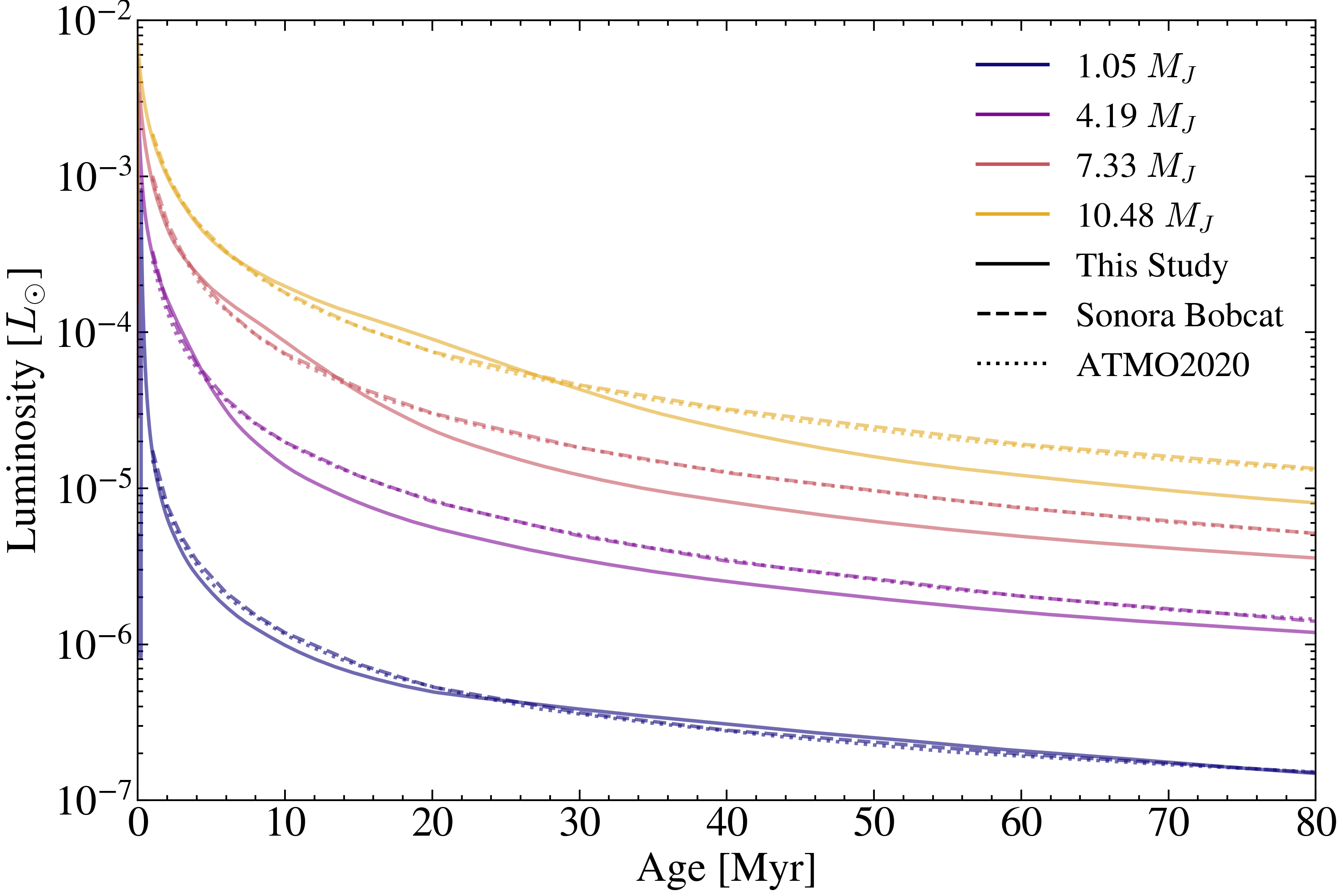}
    \caption{Comparison of our cooling curves with \textsc{Sonora Bobcat} \citep{marleySonoraBrownDwarf2021a} and \textsc{ATMO 2020} \citep{phillipsNewSetAtmosphere2020} for a range of planetary masses.}
    \label{fig:cooling_curves_comparison_other_studies}
\end{figure}

 The comparison above demonstrates that our cooling tracks are broadly consistent with established hot-start calculations. However, it does not address whether the formation mechanism itself leaves an imprint on the long-term evolution. To investigate this question, we compare our models with evolutionary tracks derived from the core accretion scenario, using the different setups presented in \citet{mordasiniLuminosityYoungJupiters2013}. These models are based on the planet formation framework originally developed by  \citet{alibertModelsGiantPlanet2005}, which was later extended to include the subsequent long-term evolution by \citet{mordasiniCharacterizationExoplanetsTheir2012, mordasiniCharacterizationExoplanetsTheir2012b}. Figure \ref{fig:cooling_curves_comparison_different_formation_mechanism} shows our cooling tracks for three planetary masses (1, 7 and \SI{12}{\MJ}) compared to  evolutionary models associated with core accretion under three different assumptions: hot accretion, cold accretion with a core mass of 20 M$_{\oplus}$ and cold accretion with a core mass of 127 M$_{\oplus}$. We find that for all planetary masses considered, the hot accretion models are very similar to our cooling tracks. 
 This suggests that luminosity alone (especially at ages beyond a few million years) cannot be used to distinguish between planets formed via disk instability and those formed via core accretion. On the other hand, the cold accretion models are rather different from our tracks. In the case of cold accretion the core mass plays an important role: for lower core masses, the cold accretion tracks converge with our models only at relatively late times. For more massive cores, however, the convergence occurs much earlier, making such models potentially indistinguishable from our disk instability tracks even at the young ages relevant for directly imaged planets. We note that our models do not include solid cores, although core formation may also occur in planets formed via disk instability \citep{helledCoreFormationGiant2008a}. 

The comparison with core accretion models should be taken with caution since the core accretion model does not predict a unique post-formation luminosity evolution.
The primordial planetary entropy and, more generally, the initial thermal state of the planet remain uncertain. 
For example, \citet{cummingPrimordialEntropyJupiter2018} showed that the primordial entropy can change significantly depending on the assumed heavy-element and gas accretion rates and the atmospheric opacity. \citet{berardoEVOLUTIONGASGIANT2017} investigated  how the planetary entropy depends on the accretion-shock boundary condition. It was shown that depending on the shock temperature, accretion rate and entropy of the planet's envelope at the onset of runaway accretion, the planet can form with a broad range of primordial entropies, including cold, warm, and hot states. 
Overall, in the core accretion models, different assumptions regarding the formation and accretion history can lead to a large range of solutions. Some of them are rather similar to those associated with formation via disk instability.  

\begin{figure}[h]
    \centering
    \includegraphics[width=\hsize]{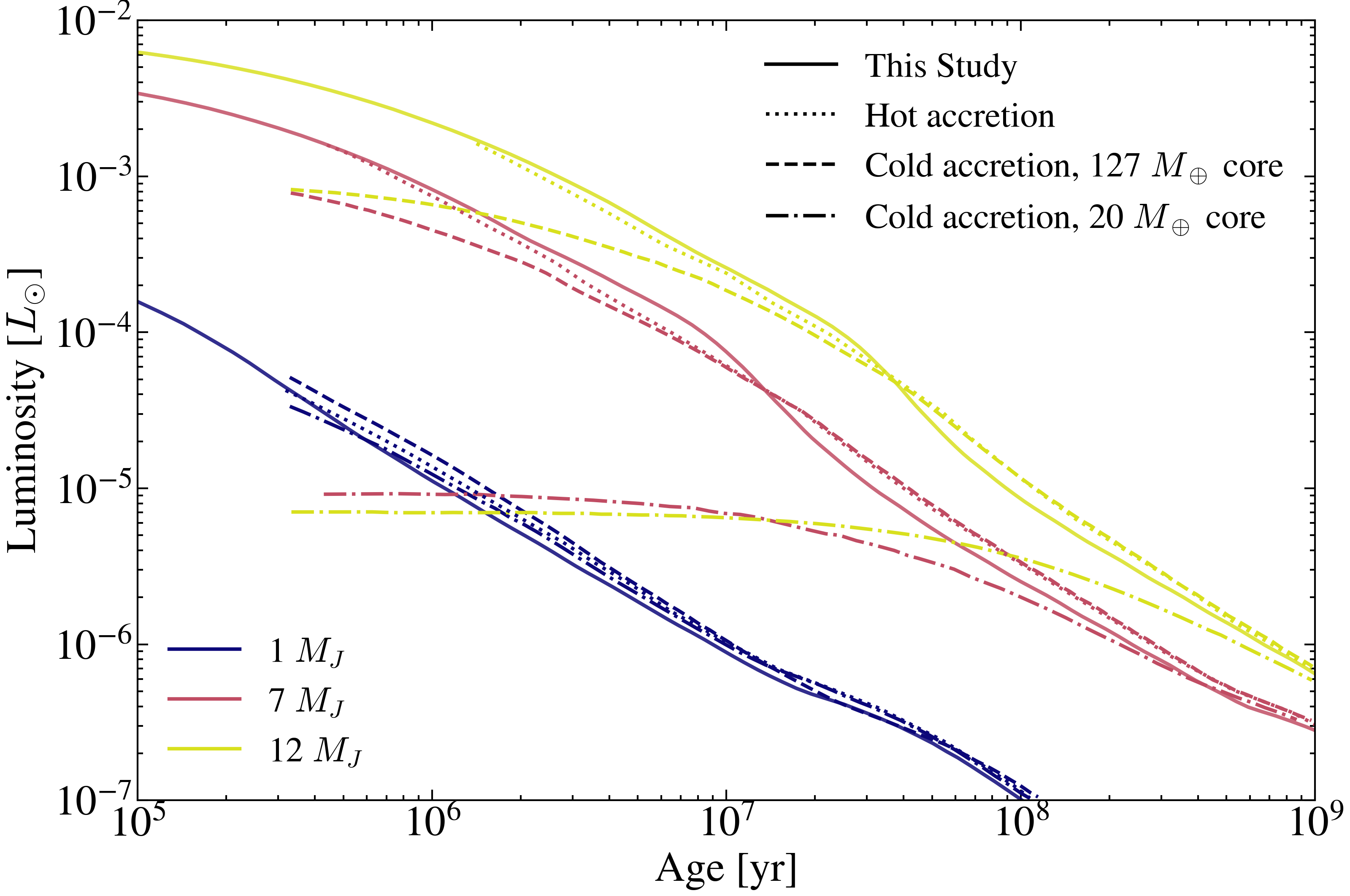}
    \caption{Comparison of our cooling tracks with cooling tracks associated with core accretion under different assumptions/outcomes from \citet{mordasiniLuminosityYoungJupiters2013}. The colours correspond to different planetary masses and the different line styles to different formation histories.}
\label{fig:cooling_curves_comparison_different_formation_mechanism}
\end{figure}

\section{Implications for the characterization of directly imaged planets} 
\label{sec:obs_with_dyn}
In this section we use our evolution models to infer the masses of planets with measured luminosities and ages.  
Directly imaged exoplanets with measured dynamical masses can be used to test the consistency of our simulations. By applying our cooling curves to planets with known bolometric luminosities and ages, we can infer their masses and compare these estimates with independently determined dynamical masses.
At present, only a small number of directly imaged exoplanets have both measured bolometric luminosities and dynamical mass constraints. This is because such planets are typically located at wide orbital separations and therefore have long orbital periods. Determining the dynamical mass requires monitoring a sufficiently large fraction of the orbit, which takes a long time for planets at large separations from their host stars.
Below, we focus on four exoplanets within the mass range considered in this work  as listed in Table \ref{tab:dyn_mass_planets}.  Figure \ref{fig:cooling_curves_with_obs_with_dyn_mass} shows the cooling curves of our evolutionary models for different masses, together with the measurements of the four exoplanets discussed below. 
\begin{figure}[h]
    \centering
    \includegraphics[width=1\linewidth]{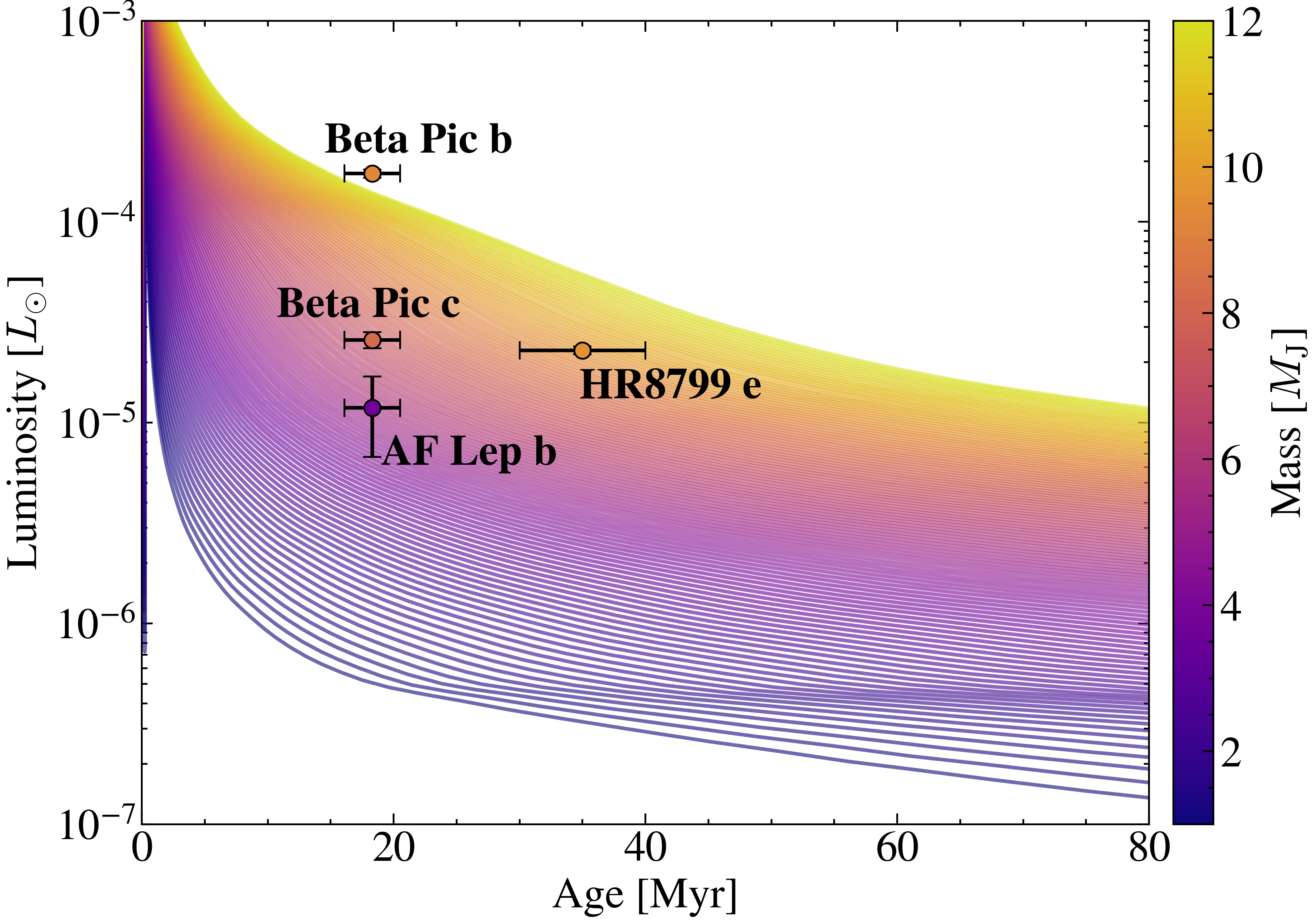}
    \caption{Cooling tracks of various planetary masses with a protostellar composition. We focus on ages typical of directly imaged planets. The data points correspond to the four exoplanets with measured bolometric luminosities and (dynamical) masses. The measured dynamical masses (mean value) are indicated by the colour of the data points.}
    \label{fig:cooling_curves_with_obs_with_dyn_mass}
\end{figure}

\begin{table*}[ht]
\centering
\renewcommand{\arraystretch}{1.3}
    \caption{Directly imaged planets with dynamical mass constraints.}
    \begin{tabular}{cccccc}
    \toprule \toprule
        Planet & Age [\SI{}{\Myr}] & log($L$/$L_\odot$) & $M_{evol}$ [\SI{}{\MJ}] & $M_{dyn}$ [\SI{}{\MJ}] & References\\ \midrule
        \textbf{HR 8799 e}       &  30 - 40 & $-4.64 \pm 0.02$ & 8.42 - 10.90 & $9.6_{-1.8}^{+1.9}$ & \makecell{\citet{faramazDetailedCharacterizationHR2021, ruffioJupiterlikeUniformMetal2026}\\ \citet{brandtFirstDynamicalMass2021}}\\
        \textbf{AF Lep b}        & $18.5_{-2.4}^{+2.0}$ & $-4.97 \pm 0.20$ & 4.11 - 6.94 & $3.75 \pm 0.5$ & \makecell{\citet{miret-roigDynamicalTracebackAge2020, grattonImplicationsDiscoveryAF2024}\\ \citet{balmerVLTIGRAVITYObservations2025}}\\
        \textbf{$\boldsymbol{\beta}$ Pic b}   & $18.5_{-2.4}^{+2.0}$ & $-3.76 \pm 0.02$ & > 11.85 & $9.3_{-2.5}^{+2.6}$ & \makecell{\citet{miret-roigDynamicalTracebackAge2020, chilcote124MmNearIR2017}\\ \citet{brandtPreciseDynamicalMasses2021}}\\
        \textbf{$\boldsymbol{\beta}$ Pic c}   & $18.5_{-2.4}^{+2.0}$ & $-4.50 \pm 0.04$ &  6.21 - 8.10 & $8.3 \pm 1.0$ & \makecell{\citet{miret-roigDynamicalTracebackAge2020, nowakDirectConfirmationRadialvelocity2020}\\ \citet{brandtPreciseDynamicalMasses2021}}\\
    \bottomrule
    \bottomrule
    \end{tabular}
    \tablefoot{This table summarizes directly imaged planets within our model mass regime that have measured bolometric luminosities, estimated system ages and dynamical mass constraints. We also report $M_{evol}$, the mass inferred from our evolutionary models using age and luminosity, enabling comparison between model-derived and dynamical mass estimates.}
    \label{tab:dyn_mass_planets}
\end{table*}

\subsection{HR 8799 e}
The first and most prominent example is the HR 8799 system, which is the first multi-planetary system discovered with direct imaging.
There are four planets in this system, HR 8799 b, c, d and e \citep{maroisDirectImagingMultiple2008, maroisImagesFourthPlanet2010}, while we currently only have the dynamical mass of planet HR 8799 e. 
Their current projected separations are 68, 38, 24 and 14 astronomical units, respectively, and the planets are expected to be in a mean motion resonance chain \citep[e.g.,][]{fabryckySTABILITYDIRECTLYIMAGED2010, konopackyASTROMETRICMONITORINGHR2016, wangDynamicalConstraintsHR2018, gozdziewskiExactGeneralizedLaplace2020}.
Hydrodynamical simulations suggest that the most robust scenario for the formation of this system is that they formed even further out in the disk and then migrated inwards until they became progressively trapped in the multiple resonances \citep{zurloOrbitalDynamicalAnalysis2022}. 
Additionally, \citet{schibDIPSYNewDisc2025} found in their population synthesis model, similar to previous radiation-hydrodynamic studies \citep[e.g.,][]{boleyTWOMODESGAS2009, zhuCHALLENGESFORMINGPLANETS2012, olivaModelingDiskFragmentation2020}, that multi-planetary systems are a possible outcome of the disk instability model as mass loading onto the protoplanetary disk from infall can drive the disk into multiple episodes of fragmentation. If some of these fragments survive, multiple planets form. This could explain why many systems we observe actually have multiple planets (e.g., HR 8799, WISPIT 2, PDS 70). It is therefore reasonable to assume that the objects in the HR 8799 system formed via disk instability. 
We use a bolometric luminosity of $log\left(L/L_\odot\right) = -4.64 \pm 0.02$ for planet HR 8799 e from \citet{ruffioJupiterlikeUniformMetal2026} who estimated this by computing 0.15 - 30 $\mu m$ low-resolution models using petitRADTRANS \citep{mollierePetitRADTRANSPythonRadiative2019}.
The age of the system is not well-determined. 
While it has been proposed that HR8799 is a kinematic member of the \SI{30}{\Myr} old Columba association \citep{doyonAgeHR87992010, zuckermanTUCANAHOROLOGIUMCOLUMBA2011}, \citet{leeDevelopmentModelsNearby2019} suggest that it is part of the $\beta$ Pic moving group with an estimated age of about \SI{20}{\Myr}. \citet{faramazDetailedCharacterizationHR2021} re-examined the kinematics of the system and concluded that HR 8799 either formed alone or in a molecular cloud complex that spawned also the Columba group, thus we estimate its age between \SIrange[]{30}{40}{\Myr}.
We compare these values with our cooling tracks for all metallicities, and find a mass range of \SIrange[]{8.42}{10.90}{\MJ} for HR 8799 e.
\citet{brandtFirstDynamicalMass2021} has estimated a dynamical mass for HR 8799 e of $9.6_{-1.8}^{+1.9}$ \SI{}{\MJ} by combining the orbits of all four planets with astrometric measurements from the Gaia Early Data Release 3.
The dynamical mass constraint compares very well with our predicted mass of HR 8799 e.

\subsection{AF Lep b}

Another planet with measured bolometric luminosity and a dynamical mass is AF Lep b.
The planet has independently been discovered by \citet{derosaDirectImagingDiscovery2023} and \citet{mesaAFLepLowestmass2023} using the VLT-SPHERE and by \citet{fransonAstrometricAccelerationsDynamical2023} using the Keck-NIRC2.
The host star is a $1.09 \pm 0.06$ Solar mass star of spectral type F8 \citep{grayContributionsNearbyStars2006}, has a super-solar metallicity with [Fe/H] = $0.29 \pm 0.03$ \citep{perdelwitzAnalysisPublicHARPS2024} and is part of the $\beta$ Pic moving group.
The age of this moving group is  debated: while \citet{bellSelfconsistentAbsoluteIsochronal2015} suggest an age of $24 \pm 3$ \SI{}{\Myr} according to isochronal age estimates, \citet{miret-roigDynamicalTracebackAge2020} infer  dynamical age estimates of $18.5_{-2.4}^{+2.0}$ \SI{}{\Myr}.
The planet has a semi-major axis of 9 au \citep{balmerVLTIGRAVITYObservations2025} and a bolometric luminosity of $log(L/L_\odot) = -4.97 \pm 0.20 $ \citep{grattonImplicationsDiscoveryAF2024}.
With its dynamical mass of $3.75 \pm \SI{0.5}{\MJ}$ it is the lowest-mass planet where both the dynamical mass and the bolometric luminosity are known \citep{balmerVLTIGRAVITYObservations2025}.
 When applying these values together with the age estimate from \citet{miret-roigDynamicalTracebackAge2020}, we infer a mass range of \SIrange[]{4.11}{6.94}{\MJ}, which is consistent with the dynamical mass estimate.
If instead the age estimate from \citet{bellSelfconsistentAbsoluteIsochronal2015} is adopted, the inferred mass increases to \SIrange[]{4.55}{8.01}{\MJ}. This range lies marginally above the dynamical mass estimate, clearly demonstrating the importance of the age estimate.

\subsection{$\beta$ Pic b and c}
Another system with directly imaged planets with dynamical mass constraints is $\beta$ Pic, with planets b and c.
This system has been of great interest already since \citet{smithCircumstellarDiskPictoris1984} directly imaged a circumstellar disk around the star $\beta$ Pic, the first time a circumstellar disk had ever been directly imaged.
It was found that the disk extends to more than 400 au from the star and speculated that elements of the disk could be associated with planet formation.
This has led to a variety of studies \citep{burrowsHSTObservationsBeta1995, kalasAsymmetriesBetaPictoris1995, mouilletPlanetInclinedOrbit1997, heapSpaceTelescopeImaging2000, augereauDynamicalModelingLarge2001, wahhajInnerRingsPictoris2003, golimowskiHubbleSpaceTelescope2006} that led to the direct imaging of $\beta$ Pic b by \citet{lagrangeProbableGiantPlanet2009}.
This planet is on an inclined orbit, coplanar with the circumstellar disk at a semi-major axis of about 10 au \citep{lagrangeProbableGiantPlanet2009}.
More recently, a second planet, $\beta$ Pic c, was detected via long-term radial velocity measurements from HARPS \citep{lagrangeEvidenceAdditionalPlanet2019} and subsequently directly confirmed by \citet{nowakDirectConfirmationRadialvelocity2020} with GRAVITY, marking the first direct detection of a planet initially discovered through radial velocity.
This planet resides even closer to the host star with a separation of only about 3 au.
By combining the astrometric measurements from GRAVITY for both planets, the long-term radial velocity measurements from HARPS and proper motion anomalies from Hipparcos–Gaia \citep{brandtFirstDynamicalMass2021, kervellaStellarSubstellarCompanions2022}, dynamical mass constraints of $9.3_{-2.5}^{+2.6}\SI{}{\MJ}$ for $\beta$ Pic b and $8.3 \pm 1.0 \ M_{J}$ for $\beta$ Pic c have been determined \citep{brandtFirstDynamicalMass2021}.
The bolometric luminosity estimate of $\beta$ Pic b is $log\left(L/L_\odot\right) = -3.76 \pm 0.02$ \citep{chilcote124MmNearIR2017} and for $\beta$ Pic c it is  $log\left(L/L_\odot\right) = -4.50 \pm 0.04$ \citep{nowakDirectConfirmationRadialvelocity2020}.
With an estimated age of the system of $18.5_{-2.4}^{+2.0}$ \SI{}{\Myr} \citep[as part of the $\beta$ Pic moving group;][]{miret-roigDynamicalTracebackAge2020}, according to our models we predict masses of \SIrange[]{6.21}{8.10}{\MJ} for $\beta$ Pic c and a mass above \SI{11.85}{\MJ} for $\beta$ Pic b.
Also these planets are both consistent with our cooling tracks. 

Overall, we find that the masses inferred from our disk instability cooling tracks are consistent with the dynamical mass estimates for all currently known directly imaged exoplanets within the mass range considered in this work. This agreement provides an important observational test of our evolutionary models and supports their use for deriving planetary masses from measured bolometric luminosities.
More generally, the consistency between the observations and our predictions demonstrates that evolutionary tracks based on the disk instability model can reproduce the observed luminosities of directly imaged giant planets. As a result, disk instability remains a viable formation mechanism for these objects and cannot be excluded. 
At the same time, the agreement with the available dynamical mass measurements indicates that cooling tracks alone are currently insufficient to distinguish between planets formed by disk instability or core accretion. Additional observational constraints are required to identify the dominant formation pathway of individual directly imaged planets.

\section{Summary and conclusions}
\label{sec:summaryandconclusion}
We present evolutionary tracks for giant planets formed via disk instability.
First, we follow the evolution of the extended, low-density gas clump until the dynamical collapse is initiated. We then determine a post-collapse configuration and model the long-term evolution of the protoplanet.
We consider protoplanets with masses ranging from \SIrange[]{1}{12}{\MJ} with metallicities between \SIrange[]{0.5}{2}{\Zp}.
We then used these cooling tracks and compared our inferred masses with dynamical mass estimates for  HR 8799 e, AF Lep b, $\beta$ Pic b, and $\beta$ Pic c. We also investigate the impact of metallicity on the planetary evolution and quantify its effect on the inferred masses from the age-luminosity relation. 
Our main conclusions can be summarized as follows:
\begin{itemize}
    \item The masses inferred from our evolutionary tracks are consistent with the  dynamical mass measurements of directly imaged exoplanets within the mass range considered (i.e., HR 8799 e, AF Lep b, $\beta$ Pic b, and $\beta$ Pic c), providing an important observational validation of our evolutionary models.
    \item As our cooling tracks, which are based on the disk instability model, are consistent with all dynamical mass measurements, disk instability remains a viable formation mechanisms for these objects.
    \item Both disk instability and core accretion can produce very similar luminosity evolution when core accretion models assume hot accretion. This implies that luminosity alone is generally insufficient to distinguish between the two formation pathways. 
    \item Our evolutionary tracks differ from core accretion models with cold accretion, but the convergence timescale depends on core mass. Cold accretion models with low-mass cores remain distinct until relatively late times (> $10^8$ yr), while models with massive cores converge much earlier and become similar to our tracks already at young ages ($\sim\!10^7$ yr).
    \item The assumed metallicity has a significant impact on inferred masses from luminosity measurements. For a given age and luminosity measurement, the inferred mass can differ by up to \SI{1.5}{\MJ} between the \SIrange[]{0.5}{2}{\Zp} models. 
    \item The effect of the metallicity changes with time. The largest differences occur at intermediate ages ($\SI{\sim30}{}-\SI{60}{\Myr}$), while at very young ($\lesssim \SI{20}{\Myr}$) and older ages ($\gtrsim \SI{60}{\Myr}$) the inferred masses are less sensitive to the assumed metallicity.
    \item Our evolutionary models reproduce the cooling behaviour predicted by established hot-start models across a wide range of masses and ages, with only modest differences at high masses that are primarily attributable to the treatment of atmospheric boundary conditions.
\end{itemize}

While our results show that the long-term evolution of giant planets cannot be used to discriminate between different formation models, evolutionary models (associated with a given formation path) remain a crucial tool for interpreting directly imaged planets. This is because specific physical assumptions in the formation model can still significantly affect the initial state of the planet and, consequently, its subsequent evolution. A clear example is the treatment of the accretion luminosity in core accretion models that leads to hot- and cold-start models which we can clearly distinguish at ages relevant for directly imaged planets. This demonstrates that evolutionary models can provide important constraints on the physical processes operating during the planetary formation. In the case of disk instability, this motivates further investigations. For example, our work did not include gas accretion, core formation, rotation and orbital migration, processes that can clearly affect the formation and evolution of the planets. A more detailed treatment of the dynamical collapse could also influence the early evolution and may leave imprints on the subsequent long-term evolution.

A more complete theoretical framework would allow us to test which physical assumptions lead to evolutionary tracks that are consistent with observed planets.
This will become increasingly important as the sample of directly imaged planets with well-constrained ages, luminosities and dynamical masses continues to grow. In particular, future dynamical mass measurements, including those expected from Gaia Data Release 4 at the end of 2026, will expand the number of benchmark planets available for comparison. Such a sample will allow more systematic tests of evolutionary models across different planetary masses, ages,  and orbital separations and may help identify which formation histories and model assumptions are most consistent with the observed population of giant exoplanets.

\begin{acknowledgements}
We thank the anonymous reviewer for valuable comments. The authors acknowledge the financial support of the Swiss National Science Foundation under grant 10.001.720. We thank Simon Müller, Gabriele Cugno, Mark Eberlein and Christian Reinhardt for valuable discussions and suggestions.
\end{acknowledgements}

\bibliographystyle{aa}
\bibliography{Clump_Evol_1}

\begin{appendix}
\onecolumn
\nolinenumbers

\section{Importance of the hydrogen-helium EoS}

\label{app:importance_of_eos}

As discussed in Section \ref{sec:intro}, we use different hydrogen-helium EoS for the pre-collapse and post-collapse phases. During the pre-collapse phase, the clumps have low temperatures and densities, for which the SCvH EoS provides the required coverage of the relevant parameter space. For the post-collapse evolution, however, we use the CMS EoS, which is more appropriate for the thermodynamic conditions of evolved giant planets and is commonly used in recent planetary evolution calculations. This change of EoS during the dynamical collapsed introduces a small inconsistency, because the same mass, composition and entropy do not correspond to exactly the same radius when different hydrogen-helium EoS are used. 
We quantify this effect in Figure \ref{fig:different_radii_from_eos} where we present different initial radii for different post-collapse EoS. The mass-radius relation shows the initial post-collapse radii inferred using the SCvH and CMS EoS for otherwise identical models, i.e. with the same mass, composition and entropy. Depending on the mass, the inferred radius can differ by up to 50\%. This is a significant difference in the immediate post-collapse configuration and therefore has to be assessed carefully.
To test whether this radius difference affects the subsequent evolution, we perform an additional comparison for a \SI{10}{\MJ} model. We consider three setups: a model using the CMS EoS with the larger initial radius inferred from the CMS EoS, a model using the CMS EoS but with the smaller radius inferred from the SCVH EoS, and a model using the SCvH EoS with the smaller radius (inferred from the SCvH EoS). The first two models isolate the effect of the initial radius at fixed EoS, whereas the latter two models isolate the effect of changing from the EoS at fixed initial radius.

The inferred radius and luminosity evolution of these cases is shown in Figure \ref{fig:different_luminosity_from_radii}. We find that the two models that use the CMS EoS start (but with different initial radii) rapidly converge. After an initial relaxation phase, their sizes and luminosities become similar within a few Myr. This shows that the difference in the initial radius introduced by the EoS transition is short-lived and does not affect the long-term evolution \citep[e.g.,][]{fortneyInteriorStructureComposition2010, marleauConstrainingInitialEntropy2014}.

In contrast, models that start from the same initial radius but use different hydrogen-helium EoS show a clear difference in their subsequent long-term evolution. This demonstrates that the choice of EoS itself has a significant impact on the planetary contraction and cooling, whereas the initial radius mismatch associated with switching from SCvH to CMS does not. Therefore, although the transition from SCvH to CMS introduces an inconsistency in the immediate post-collapse radius, this inconsistency does not control the long-term results. The use of the CMS EoS for the post-collapse phase is thus well motivated, because the long-term evolution is determined primarily by  the adopted hydrogen-helium EoS rather than by the transient adjustment of the initial radius \citep[e.g.,][]{militzerINITIOEQUATIONSTATE2013, chabrierNewEquationState2019, howardAccountingNonidealMixing2023, howardGiantExoplanetComposition2025}. 

\begin{figure}[h!]
    \centering
    \includegraphics[width=0.5\linewidth]{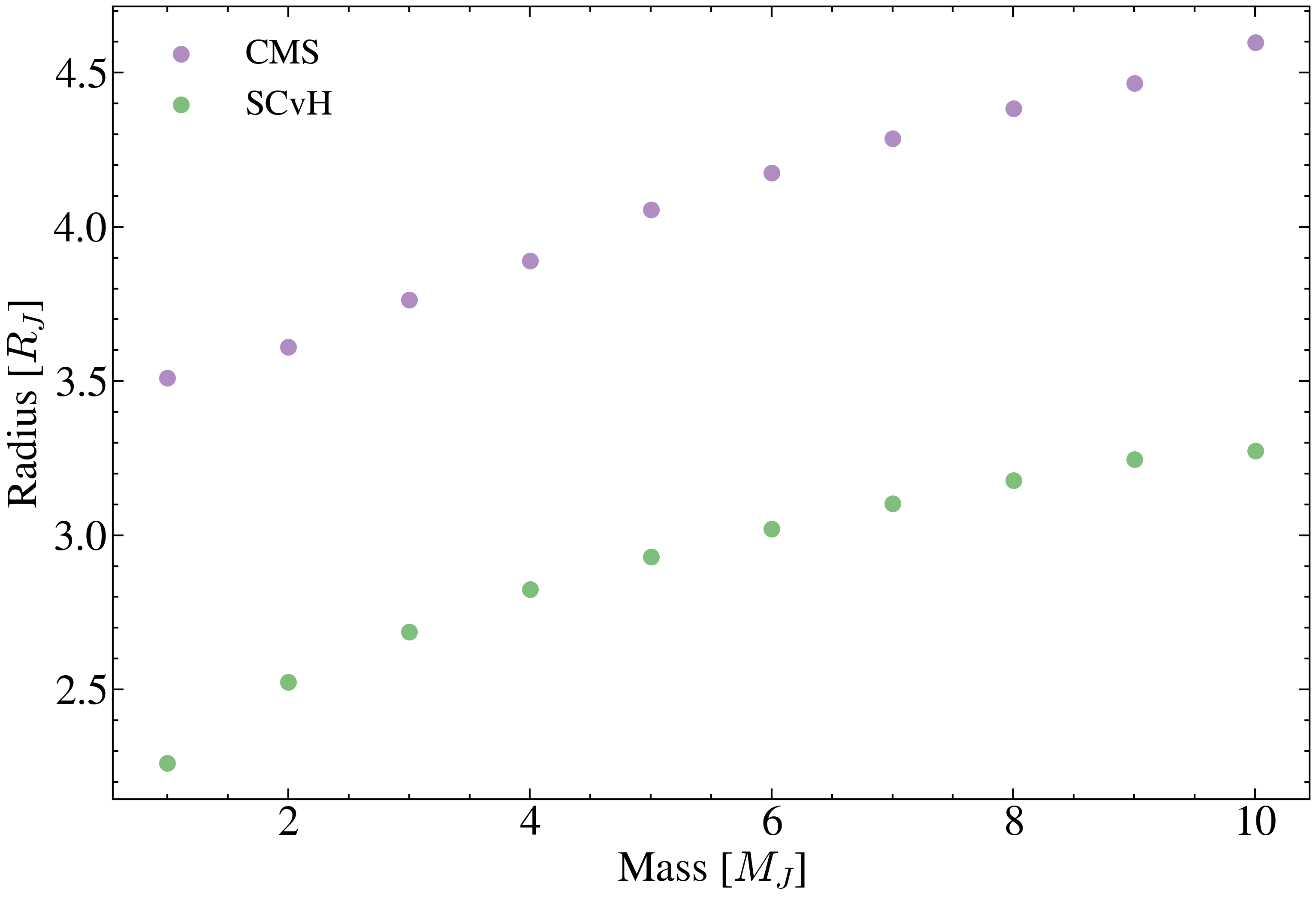}
    \caption{Different initial radii for different post-collapse EoS. After the dynamical collapse, the initial configuration of the protoplanet is determined through the total entropy of the object at the end of the pre-collapse phase and the EoS. The initial radius varies significantly depending on the chosen EoS.}
    \label{fig:different_radii_from_eos}
\end{figure}

\begin{figure}[h!]
    \centering
    \includegraphics[width=\linewidth]{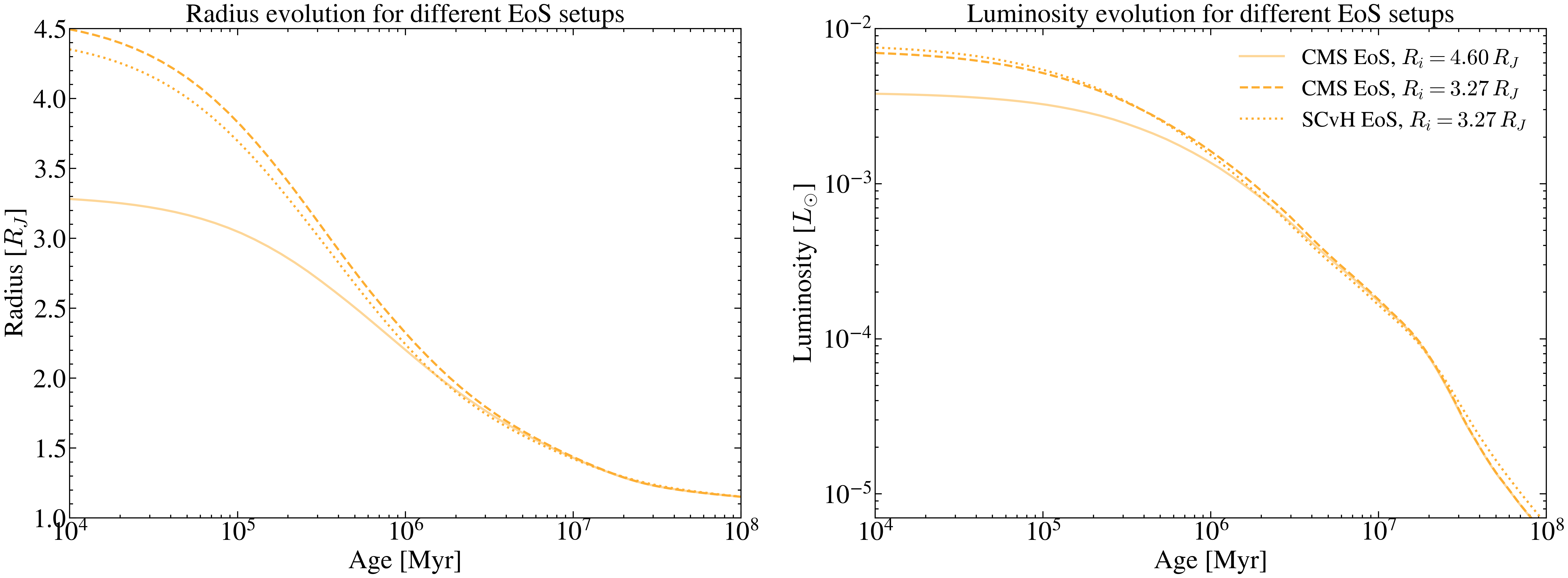}
    \caption{The effect of the initial radius and the used EoS on the long-term evolution. We show cooling tracks of a \SI{10}{\MJ} protoplanet for different EoS and initial radii. The solid and dashed lines both us the CMS EoS but with vastly different initial radii, illustrating whether such a difference remains visible on relevant timescales. The dotted line uses the SCvH EoS with the same initial radius as the dashed CMS track, allowing us to investigate the influence of the EoS choice itself. At timescales relevant for direct imaging observations the choice of EoS is much more relevant.}
    \label{fig:different_luminosity_from_radii}
\end{figure}

\newpage

\section{Additional Comparisons}
\subsection{Comparison with the \textsc{sonora red diamondback} models}
\label{sec:comparison_diamondback}

As discussed in the main text, our simulations do not include clouds, condensation or atmospheric chemistry which could affect the planetary cooling. As a result, in Section  
\ref{sec:comparison} we compared our results only to cloud-free models.  
Here, we compare our models to the \textsc{sonora red diamondback} models \citep{morleySonoraSubstellarAtmosphere2024, davisSonoraSubstellarAtmosphere2025} that have a more realistic atmosphere. They employ a hybrid model that includes condensation and cloud formation at early times, while the object is still hot. After the L/T transition, which occurs at an effective temperature of about \SI{1300}{\K} and marks the shift in spectral classification from L-type to T-type, they transition to a cloud-free model \citep[see][for details]{morleySonoraSubstellarAtmosphere2024}. The results for different planetary masses are shown in Figure \ref{fig:cooling_curves_comparison_diamondback}.  
We find that at very young ages the luminosities are similar probably since under these conditions cloud formation and condensation are negligible. Once the planets cool to luminosities of a few $\sim4 \times 10^{-4}$ L$_{\odot}$ \textsc{sonora red diamondback} models cool slightly faster. 
This is the case until the L/T transition at $\sim1 \times 10^{-4}$ L$_{\odot}$, when the dusty clouds clear and the cooling stalls. As the models converge to cloud-free models, the \textsc{sonora red diamondback} models become comparable to our models again. For the \SI{2.1}{\MJ} model the effective temperature is below \SI{1300}{\K}, thus it skips this dusty cloud phase entirely. The remaining differences are expected to be caused by the different treatment of the atmospheric boundary condition \citep[see][and references therein for further discussion]{morleySonoraSubstellarAtmosphere2024, davisSonoraSubstellarAtmosphere2025}.

\begin{figure}[h!]
    \centering
    \includegraphics[width=0.5\linewidth]{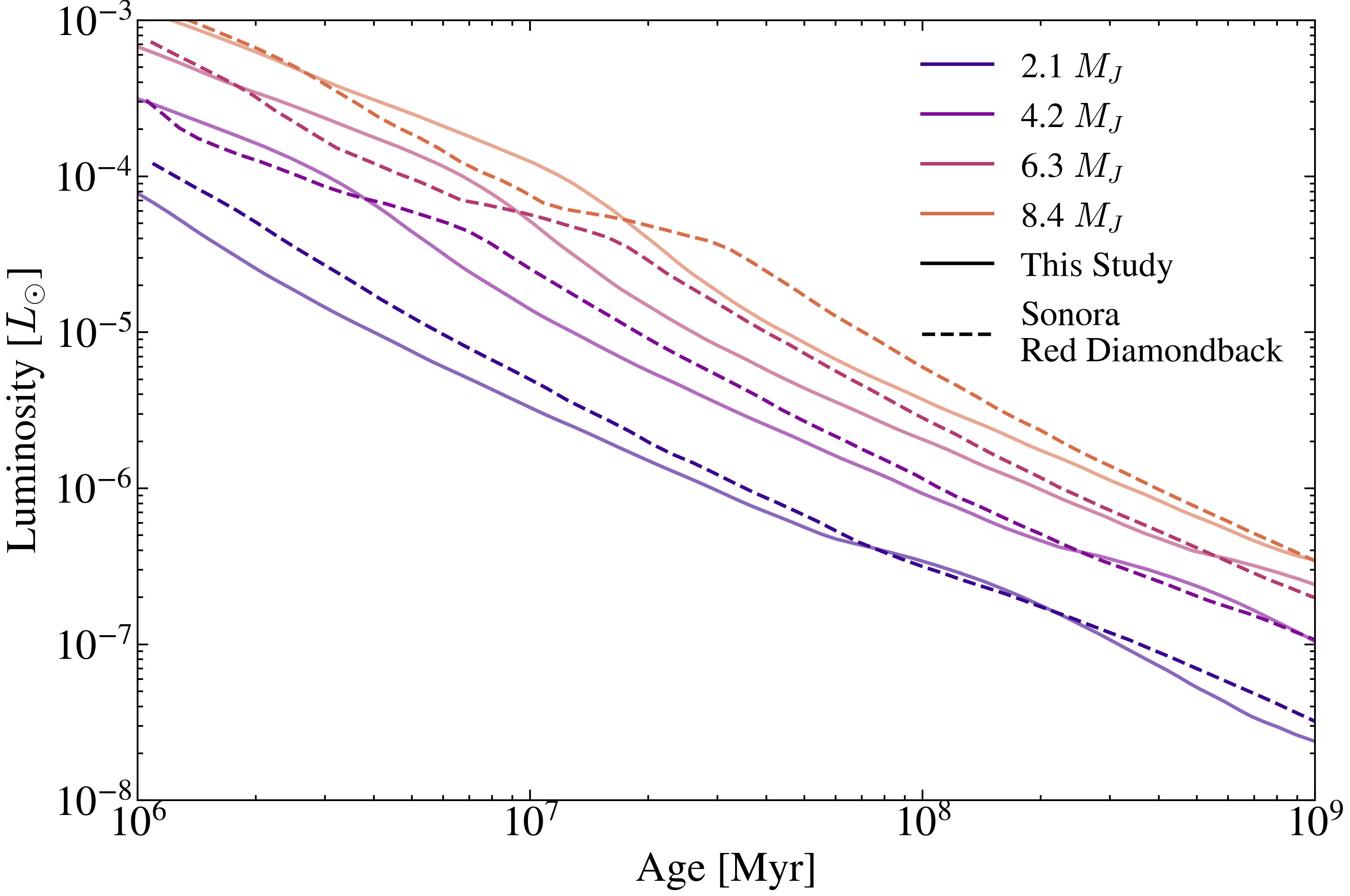}
    \caption{A comparison of our luminosity evolution with the ones inferred with \textsc{sonora red diamondback} \citep{morleySonoraSubstellarAtmosphere2024, davisSonoraSubstellarAtmosphere2025} for several planetary masses.}
\label{fig:cooling_curves_comparison_diamondback}
\end{figure}

\subsection{The planetary Hertzsprung–Russell diagram}
Recently, \citet{gottsteinPlanetaryFormationTracks2026} presented the Hertzsprung–Russell diagram (HRD\footnote{The HRD is a fundamental stellar-evolution diagram that relates a star's luminosity to its effective temperature, revealing distinct populations and evolutionary stages.}) expected from planet formation within the Bern model framework for both pebble and planetesimal accretion. For comparison, Figure \ref{fig:HRD} presents the inferred HRD (luminosity vs.~effective temperature) from our models for planets with masses between 1 and 12 M$_J$ for a proto-solar 
composition. The apparent irregularity at the onset of the pre-collapse evolutionary tracks arises from the delicate process of constructing the initial models, as described in Section \ref{sec:methods}.

\begin{figure}[h]
    \centering
    \includegraphics[width=0.5\linewidth]{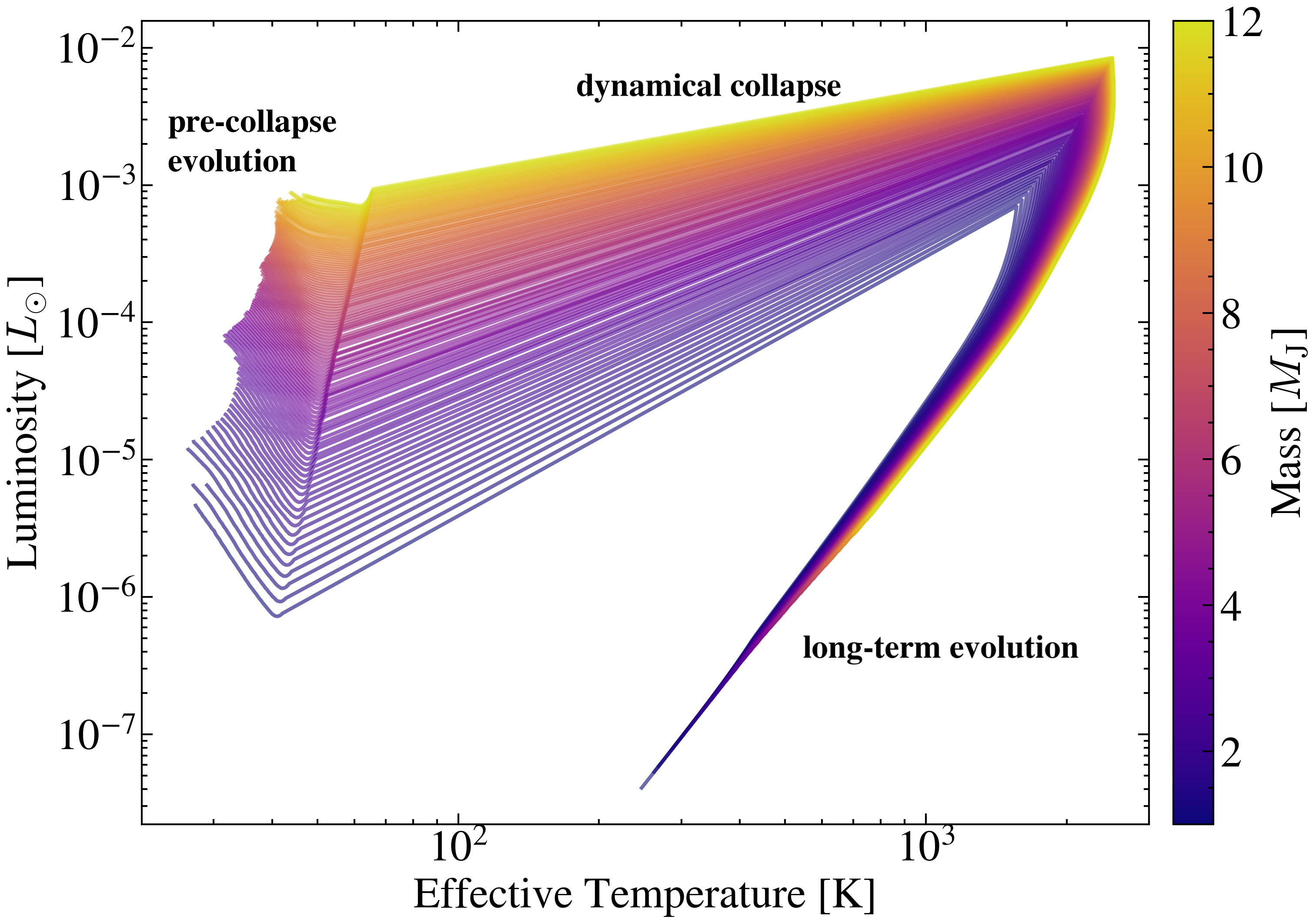}
    \caption{The planetary Hertzsprung–Russell diagram (HRD) for different planetary masses with a proto-solar 
composition. Data is shown for the formation phases and until 100 Myr of the evolution phase, rather than for the full duration as presented in Fig. \ref{fig:cooling_curves}. Also, note that the dynamical collapse corresponds to very short timescales (only a few years)}.
    \label{fig:HRD}
\end{figure}

\end{appendix}
\end{document}